\documentclass[12pt]{article}        

\usepackage{amsmath}
\usepackage{amssymb}
\usepackage{euscript}

\usepackage{epsfig}

\usepackage{cite}

\usepackage{alltt}

\usepackage{epic}
\usepackage{fancybox}

\usepackage{html}

\def\Order#1{${\cal O}(#1$)}
\def\lint{\int\limits}
\def\bbeta{\bar{\beta}}

\newcommand{\Keu}{\EuScript{K}}

\newcommand{\Reu}{\EuScript{R}}

\newcommand{\Rcal}{{\cal R}}

\begin{document}                     

\allowdisplaybreaks

\begin{titlepage}

\begin{flushright}
{\bf  CERN-TH/98-242  
\\ UTHEP-98-0702}
\end{flushright}

\vspace{2mm}
\begin{center}
{\bf\LARGE
Monte Carlo Program KoralW 1.42\\
for All Four-Fermion Final States\\
in $e^+e^-$ Collisions$^\dag$
}
\end{center}

\vspace{2mm}
\begin{center}
  {\bf   S. Jadach$^{a,b}$,}
  {\bf   W. P\l{}aczek$^{e,a}$,}
  {\bf   M. Skrzypek$^{a,b,\ddag}$,}
  {\bf   B.F.L. Ward$^{a,c,d}$}
  {\em and}
  {\bf   Z. W\c{a}s$^{b,a,\ddag}$ }
\\
\vspace{1mm}
{\em $^a$CERN, Theory Division, CH-1211 Geneva 23, Switzerland,}\\
{\em $^b$Institute of Nuclear Physics,
  ul. Kawiory 26a, 30-055 Cracow, Poland,}\\
{\em $^c$Department of Physics and Astronomy,\\
  The University of Tennessee, Knoxville, Tennessee 37996-1200,}\\
{\em $^d$SLAC, Stanford University, Stanford, California 94309,}\\
{\em $^e$  Institute of Computer Science, Jagellonian University,\\
        ul. Nawojki 11, Cracow, Poland}
\end{center}

\vspace{2mm}
\begin{abstract}
The Monte Carlo program KoralW version 1.42 is presented.
It generates all four-fermion final states with multibranch
dedicated Monte Carlo pre-samplers and complete, massive, Born matrix elements.
The presamplers cover the entire phase space. Multiphoton 
bremsstrahlung is implemented in the ISR approximation within the YFS 
formulation with the ${\cal O}(\alpha^3)$ leading-log matrix element. 
The anomalous $WWV$ couplings are implemented in CC03 approximation. 
The standard decay libraries (JETSET, PHOTOS, TAUOLA) are interfaced. 
The semi-analytical CC03-type code KorWan for differential and total 
cross-sections is included.
\end{abstract}

\begin{center}
{\it Computer Physics Communications} {\bf 119} (1999) 1
\end{center}

\vspace{2mm}
\footnoterule
\noindent
{\footnotesize
\begin{itemize}
\item[${\dag}$]
Work supported in part by 
Polish Government grants 
KBN 2P03B08414, 
KBN 2P03B14715, 
the US DoE contracts DE-FG05-91ER40627 and DE-AC03-76SF00515,
the Maria Sk\l{}odowska-Curie Joint Fund II PAA/DOE-97-316,
and the Polish-French Collaboration within IN2P3 through LAPP Annecy.
\item[${\ddag}$]
Supported at the time of performing part of this study by 
a stipend within the EU grant no ERBCIPDCT940016 at the
{\em Inst. f\"ur Theoretische Physik, Karlsruhe Universit\"at, Karlsruhe,
Germany.}
\end{itemize}
}

\vspace{1mm}
\begin{flushleft}
{\bf CERN-TH/98-242 
\\ UTHEP-98-0702
\\ July 1998}
\end{flushleft}

\end{titlepage}

\noindent{\bf NEW VERSION SUMMARY}
\vspace{10pt}

\noindent{\sl Title of the program:} KoralW, version 1.42.

\noindent{\sl Reference to original program:}
Comput. Phys. Commun. {\bf 94} (1996) 215.

\noindent{\sl Authors of original program:}
M. Skrzypek, S. Jadach, W. P\l{}aczek and Z. W\c{a}s

\noindent{\sl Computer:} 
any computer with the FORTRAN 77 compiler and the UNIX operating system

\noindent{\sl Operating system:}
UNIX (program tested under AIX 4.x, HP-UX 10.x), Linux

\noindent{\sl Programming language used:}
FORTRAN 77

\noindent{\sl High-speed storage required:}  $<$ 10 MB



\noindent{\sl No. of cards in combined program and test deck:}
about 30,000 plus 11,261+5,970+2,264 of physics generators libraries 
(JETSET+TAUOLA+PHOTOS) plus 369,331 of an external matrix element library (GRACE).


\noindent{\sl Nature of the physical problem:}
The $W$-pair production and decay is and will be
used as an important data point for precise tests
of the standard electroweak theory at LEP2 and higher energies.
The effects due to background processes, QED bremsstrahlung 
and apparatus efficiency have to be subtracted from the data.
The program deals with all $e^+e^-$ processes leading to 
4-fermion final states accompanied with multiphoton
initial-state radiation.
It also includes the effects of 
the Coulomb correction, `naive' QCD, anomalous couplings,
quarks hadronization, $\tau$ decays, and photon radiation in leptonic decays.

\noindent{\sl Method of solution:}
The Monte Carlo methods are used to simulate all 4-fermion final-state
processes in the $e^+e^-$ collisions in the presence of multiphoton
initial-state radiation. The latter is described in the framework
of the YFS exclusive exponentiation.
The $W$-pair production is included in a `natural' way as a subset
of the Feynman diagrams for the above processes, but it can also
be generated exclusively by switching to the so-called CC03 process.
The Monte Carlo generation is done on an event-by-event basis,
with constant or variable weights, where
an event is represented by flavours and four-momenta
of all respective particles -- supplemented with a collection
of weights, if the variable weight option is chosen.   
After the event generation is completed the program provides
the cross sections together with their statistical errors
for all the processes involved.
Any experimental cuts  and apparatus efficiencies may be introduced
easily by rejecting some of the generated events.

\noindent{\sl Restrictions on the complexity of the problem:}
Only processes with $4$-$fermion$ final states are considered.
QED radiative corrections are implemented in terms of multiphoton ISR
in the YFS Monte Carlo framework with the ${\cal O}(\alpha^3)$ LL-type
matrix element. For the CC03 subset of diagrams the Coulomb correction
for the intermediate $WW$ states is also included. 
The final-state QED radiation is generated for charged leptons with
the help of the program PHOTOS in the LL approximation (up to two photons).
QCD effects are included in the so-called ``naive QCD'' approximation.
A part of electroweak corrections is incorporated in the ``improved
Born approximation'' (through appropriate renormalization scheme).
Anomalous triple gauge boson couplings are implemented only in the
CC03 subset of diagrams (i.e. the $W$-pair production). 
Quadruple gauge boson couplings are not implemented.
The $\tau$-decays and quark hadronization are performed, respectively, 
with the help of the dedicated packages TAUOLA and JETSET.

\noindent{\sl Typical running time:}
On IBM PowerPC M43P240 (266 MHz, 65 CERN units) installation one needs:
(a) 
2.5 sec per 1000 constant-weight events 
and 0.6 sec per 1000 variable-weight events for CC03 matrix element and
(b)
12500 sec per 1000 constant-weight events 
and 6 sec per 1000 variable-weight events
for a complete 4-fermion matrix element (GRACE).
These results are for a {\em default/recommended} 
setting of input parameters but with {\em all} decay libraries switched
OFF. 


\newpage

\section{Introduction}

As LEP2 approaches the 200 GeV centre-of-mass 
energy regime, the need for reliable
calculations of the so-called signal CC03 and general background processes in
$e^+e^-\rightarrow W^+W^-\to f_1\bar{f}_2f_3\bar{f}_4$ 
becomes more and more immediate. 
Moreover, with the added need for the accommodation of arbitrary
detector cuts in these calculations, the only practical solution is the
Monte Carlo event generator realization of the calculation, wherein these
cuts may be imposed on an event-by-event basis. In the present paper, we
provide, in version 1.42 of the program KoralW~\cite{koralw:1995a} such an
event generator in which all relevant $4$-$fermion$ processes, 
both charged-current (CC) and neutral-current (NC) ones,
in $e^+e^-\rightarrow f_1\bar{f}_2f_3\bar{f}_4$,  
in the LEP2 200 GeV regime are realized
in the presence of the YFS~\cite{yfs:1961} exponentiated initial-state 
multiple photon radiation (ISR).

More precisely, in the $WW$ sector this version of the program 
has as its ultimate objective
the sub-per-cent precision regime of $0.5\%$ as called for in the 
LEP2 Workshop Report of Ref.~\cite{yr:lep2}. Thus, when compared
with version 1.02~\cite{koralw:1995a}, the inclusion
of the $4$-$fermion$ background processes for the
$e^+e^-\rightarrow W^+W^-\to f_1\bar{f}_2f_3\bar{f}_4$ 
signal is an essential improvement. 
This is described in detail in Section 3. In addition,
new physics anomalous couplings are now featured for the CC03 class of
graphs -- these were absent from version 1.02. The ISR is now calculated
through the ${\cal O}(\alpha^3)$ LL in the YFS
exponentiated framework; in Ref.~\cite{koralw:1995a}, it was only calculated
through the ${\cal O}(\alpha^2)$ LL in the same framework.
The Coulomb correction is now implemented in a way that is
more reliable near the threshold, following the authors of 
Ref.~\cite{khoze:1995}.
The leading QCD correction, as well as the CKM matrix, is 
now featured following Ref.~\cite{WW-YR:1996};
in version 1.02, these corrections are absent. Two other effects featured
in the current version, which are missing in version 1.02, are the
colour reconnection effect, which we model after Ref.~\cite{Wmass-YR:1996},
and the 
Bose-Einstein effect, which we take after 
the authors of Ref.~\cite{JadachZalewski:1997}. 
In addition, three choices of
renormalization scheme are available.
Considerable technical precision checks of the program have been made,
using semi-analytical results as described below, so that we have
established the technical precision of the program at the level of
$0.2\%$. 
The primary missing ingredient for the final step to a total
precision tag of $~0.5\%$ is the implementation of the 
exact ${\cal O}(\alpha)$ electroweak corrections for the CC03 class
of graphs in the YFS exponentiated framework, with the consequent
intermediate-state $n(\gamma)$ soft radiative effects, as we have
already published using the program YFSWW3-1.11 in Ref.~\cite{placzek:1998}.
These effects will be incorporated in a later version of 
KoralW~\cite{jadach:preparation}. As a result of the studies done in 
Refs.~\cite{koralw:1995b,koralw:1997,placzek:1998} 
and for the preparation the current paper,
we can already safely set, in the current version 1.42 of KoralW,
the precision tag for the cross section normalization, for the process 
$e^+e^-\rightarrow W^+W^-\to f_1\bar{f}_2f_3\bar{f}_4$, 
to $2\%$ in the LEP2 energy regime, where it is understood that
a cut of some type is used to define the two $W$'s.
The same $2\%$ precision tag should remain valid also for the $Z$-$Z$ physics, 
even though a formal study is missing. 
This, however, may not be true in 
general $4$-$fermion$ final-state processes. Here, the precision may depend
very much on a particular final state under consideration, a choice of
physical observables, experimental cuts, events selection criteria, etc. 
Thus, it requires a more dedicated study. The most problematic
are the processes with electron(s) and/or positron(s) in the final state,
particularly 
when regions of small scattering angles are allowed (e.g. selection criteria
allow some of the particles to be lost in the beam pipe) and/or events with 
high-$p_T$ photons are accepted.   
Various problems that can be encountered there are discussed in some detail
in Subsections \ref{CUTS} and \ref{NONISR}.
Nevertheless, at the Born level, our program can be used in all regions of
the phase space, including the most singular ones. 
Also, an important technical step has been achieved,  which is indispensable
for a reliable solution of bremsstrahlung implementation in the general case.

Let us also point out that contributions of the $4$-$fermion$ processes
from some regions of the phase space are implicitly included in 
$2$-$fermion$ generators as a part of radiative corrections (analytically
cancelled out with the leading ${\cal O}(\alpha^2)$ virtual corrections).
One needs to be cautious about this point so as to avoid double counting.

The outline of the paper is as follows. We present in detail the overall
structure of the program in Section 2. In Section 3, we fully describe 
the 4-fermion phase-space generation and the corresponding matrix elements
used in the calculations of the program, the generation of the
energy distribution due to ISR in the program, and the semi-analytical
CC03 distributions contained in the program. Section 4 describes the
practical use of the program, so that it should help the
user to take advantage of the program's capabilities in an efficient and
easy manner. Appendices contain useful technical information on the
construction and use of the program, its matrix elements, its input/output, 
etc.

\section{Structure of the Program}

In this section we provide the reader with a brief guide of the KoralW
program. We will describe its main routines,
libraries and interfaces.
We want to note here that the program, in its distribution version, is
prepared for a UNIX-type operating system that supports directories and
{\tt make} command. Namely, we have divided the source code
into a number of subdirectories, in order to make the structure more
transparent and easier to handle. Also the structure of {\tt
Makefile}-s is 
provided, for easier compilation, as well as some other auxiliary functions
(e.g. clean-up). This structure, which is in fact not very complicated, 
can be in principle avoided and code can be transformed into a single
FORTRAN file. 
Some care must be taken  while handling the {\tt ampli4f.grc.all} 
routines where certain {\tt include} files with the same name, have different 
contents in different subdirectories. Also, note that the following three
directories: {\tt ampli4f}, {\tt ampli4f.grc.all} and 
{\tt ampli4f.grc.all-old} are mutually exclusive, i.e.\ they contain three
versions of the same routines and only one at a time can be used by
the program.

\subsection{{\tt ampli4f}---Template of the Library}

Dummy directory for the external matrix element library, to be filled in
by the user. The front-end routines {\tt ampini} and {\tt amp4f} are
expected by KoralW to be put here by the user. We expect the user to replace
this directory with his/her own code, e.g. with their favourite 
parametrization of anomalous couplings.

\subsection{{\tt ampli4f.grc.all}---All Four-Fermion Library}

An implemented library of external matrix elements for all four-fermion
final states has been generated by the GRACE v.~2 package \cite{GRACE2}.
This dedicated code has been provided for KoralW
with the courtesy of the Minami-Tateya Group of KEK.

The directory {\tt amp4f} contains routines for the actual computation
of matrix elements. The auxiliary library is located in the directory
{\tt channel}. 
The routines specific to KoralW, which transmit and reset
parameters and dip-switches of GRACE routines are located in the
directory {\tt grc4f\_init}.
Finally, the front-end routines {\tt ampini} and {\tt amp4f}  
are in the {\tt amp4f} directory.

\subsection{{\tt ampli4f.grc.all-old}---Old CC-all Library}

An old version%
\footnote{See Sect.\ \ref{old-grace} for explanation why we have decided
 to keep this ``old'' package.
}
 of the library of external matrix elements for all $WW$-type
final states, generated by the GRACE v.~1 package \cite{GRACE1}.
This library contains all the scripts capable of building the FORTRAN
code from scratch with the help of the symbolic package GRACE. The
scripts are located in the subdirectory {\tt grace4f\_init}. All of the
FORTRAN files and directories created by the scripts can be identified by
the word expression ``{\tt .auto.}'' in their names. 
Similarly, the template files used by
the scripts are marked with the word expression ``{\tt .template.}''.

\subsection{{\tt demo}---Demonstration Programs}
This subdirectory contains two demonstration subprograms: 
{\tt KWdemo} and {\tt KWdemo2}.
Generally the user is supposed to provide his/her own main program.
Nevertheless the two subprograms quoted here are simple examples
of a main program.

{\tt KWdemo} is the first of them.
It has a double role 
as a useful template 
and as a first cross-check that the MC generator KoralW 
runs correctly on a given installation.
The essential part of this program is a loop in which a series of KoralW 
events is generated.
It also reads the input from a disk file, but no histogramming is performed
and most of output is from the generator itself.  
At the end of the program, a MC integrated
cross section of KoralW is compared with a semi-analytical result from KorWan.
The program is compiled/linked and executed, with the help of the makefile, for
two data sets, as follows: {\tt make KWdemoCC03} and {\tt make KWdemoGRCall}.
The two data sets for the two separate runs are:
\begin{itemize}
\item
  {\tt demo.14x/190gev/KWdemo.input.CC03}
  for which the CC03 ``signal'' process is simulated 
  (internal matrix element), the ISR is on,
  with hadronization, tau decays, no CKM mixing, weighted events.
\item
  {\tt KWdemo.input.GRCall}
  for all four-fermion matrix elements form GRACE (external ME),
  hadronization, tau decays, weighted events.
\end{itemize}

{\tt KWdemo2} is the second example.
Apart from the loop in which series of KoralW events is produced
the example of {\em histogramming} with proper normalization is also included.
It is done in the subprogram {\tt ROBOL}.
The program is compiled/linked and executed, with the help of the makefile, for
two data sets, as follows: {\tt make KWdemo2HADR} and {\tt make KWdemo2SEMI}.
The two data sets provided for the two separate runs are:
\begin{itemize}
\item
  {\tt KWdemo2.input.HADR}
  CC03 ``signal'' process is simulated, (internal matrix element), 
  the ISR is on, all four-quark channels only -- selected with 
  {\tt Umask} matrix (see Appendix B for a definition),
  {\tt WtMain} = 1 events.
\item
  {\tt KWdemo2.input.SEMI}
  CC03 ``signal'' process is simulated (internal matrix element), 
  the ISR is on,
  anomalous triple gauge boson coupling constants are activated, see data file,
  semi-leptonic channel only -- selected with the {\tt Umask} matrix.
  Variable-weight events are generated and analysed.
\end{itemize}

\subsection{{\tt glib}---Histogramming}
A handy FORTRAN histogramming package, GlibK \cite{glibk:1995}, is provided in 
this directory. It is used by KoralW both for hard-coded internal 
bookkeeping and for some optional ``external''  tests. 
The package is similar in
its usage to the  classic HBOOK package of CERNLIB.

\subsection{{\tt interfaces}---Interfaces to the Libraries}

Interfaces to external libraries are collected in this directory.
\begin{enumerate}
\item
  {\tt tohep} fills in the {\tt /HEPEVT/} common block, 
  decays $\tau$ leptons (with TAUOLA)
  and generates bremsstrahlung in $W$ decays as well as secondary decays 
  (with PHOTOS).
\item
  {\tt tohad} does the hadronization (with the help of JETSET) along with 
  necessary colour (re)connection.
\item
  The interface to the external matrix elements is provided by the routines
  {\tt ampinw} and {\tt ampext}. The first performs the necessary
  initializations. It calls the user-supplied initialization routine 
  {\tt ampini}. 
  Some additional parameters (apart from {\tt xpar}) 
  can be reached by the user with the help of {\tt masow} and 
  {\tt kwparm2} routines.
  The {\tt ampext} transmits to KoralW the value of the external matrix
  element that it calculates by a
  call to the user-supplied routine {\tt amp4f}. It also allows, with the help
  of the dip-switch {\tt key\_cms\_eff}, choosing the four-vectors supplied 
  for this calculation to be in the LAB or effective CMS (default) frames. 
\end{enumerate}

\subsection{{\tt kwlund}---JETSET and PHOTOS libraries}
JETSET v. 7.4 \cite{jetset:1987} and PHOTOS \cite{photos:1994} 
are located here.
Note that the common block {\tt /HEPEVT/} is expected to contain
single-precision ({\tt REAL*4}) variables.

\subsection{{\tt model}---Matrix Elements}
The ``model'' weights are calculated here.
\begin{enumerate}
\item
  The CC03 matrix element is calculated by {\tt wwborn}. It takes as an
  input massive four-vectors of final fermions (assumed to be in their
  rest frame with beams along the $z$-axis, $e^-$ in the +$z$ direction). 
  Next, it
  ``reduces'' these four-vectors to the massless ones\footnote{
                  Note that for the CC-all and NC-all modes the fully 
                  massive matrix element is used, see Sect.~\ref{4flib}.}  
  in three ways.
  The ``sophisticated'' method, which reconstructs angles ({\tt invkin}) and
  rebuilds four-vectors with the same angles and zero masses 
  ({\tt kineww}). The next method simply rescales the three-momenta, thus 
  breaking the overall momentum conservation. The third provided option is 
  ``no reduction at all''. These massless four-vectors are transmitted to
  {\tt wwborn\_massless} that does the actual calculation with the help of
  the {\tt wwprod} and {\tt wdecay} routines. If anomalous couplings are
  requested the {\tt wwamgc} routine is called instead of {\tt wwprod}.
\item
  The ISR photonic corrections (the so-called beta-functions) up to the third
  order are provided by the routine {\tt betar}, which in turn calls the
  {\tt d\_isr*} routines to construct the actual real and virtual
  bremsstrahlung contributions.
\item
  The Coulomb correction is located in {\tt culmc} routine. It is used by
  Monte Carlo routines. Semi-analytical routines have an identical
  correction implemented in the {\tt culsan} routine in {\tt semian}
  directory. 
\end{enumerate}

\subsection{{\tt semian}---Semi-analytical Routines}
This directory contains a package for semi-analytical calculations.
\begin{enumerate}
\item
  The total cross section $\sigma$, with or without ISR, is provided by 
  the {\tt korwan} routine. 
  For the differential $d\sigma/d\log v$ it uses the {\tt yfspho}
  function. For the total $\sigma$ with ISR, {\tt korwan} 
  integrates {\tt yfspho} and,
  in the case of no radiation, $\sigma$ is provided by the function {\tt
    xsmuta}; this in turn integrates the {\tt d1muta}, which
  integrates {\tt d2muta}, the actual two-dimensional differential
  cross-section. The function {\tt yfspho} provides various kinds of 
  one-dimensional 
  photonic distributions  $d\sigma/dv$ (with soft residual subtracted) 
  or $d\sigma/d\log v$.
  It uses {\tt xsmuta} as well.
\item
  The one-dimensional distribution of the single-$W$ invariant mass is
  provided by the {\tt s1wan} function.
\item
  The two-dimensional distribution of the double-$W$ invariant mass is
  provided by the {\tt s1s2wan} function.
\item
  The average mass, 
  $(1/\sigma)\int dvds_1ds_2 (\sqrt{s_1}+\sqrt{s_2}-2M_W)d\sigma/(dvds_1ds_2)$,
  is calculated by the {\tt mavrg} routine with the help of {\tt korwan} with
  negative $s$-variable input parameter.
\item
  The average mass loss, $(\sqrt{s}/2)(1/\sigma)\int dv\; v\; d\sigma/dv$, is
  calculated by the {\tt mloss} routine with the help of {\tt korwan} with
  negative {\tt keypho} input parameter.
\end{enumerate}

\subsection{{\tt tauola}---TAUOLA  library}
It is the directory with the TAUOLA library \cite{tauola:1993}
for simulation of $\tau$-decays.
Note that, as usual, in the distribution version of TAUOLA the parameters 
in the $\tau$-decay modes are not adjusted appropriately. 
We recommend that user replaces this version of TAUOLA by the version
of his/her own collaboration\footnote{
                                      In case it is impossible, please
                                      contact the authors.}.
   
\subsection{{\tt korww}---``Central Processing Unit''}
\label{korww}

This directory contains the actual Monte Carlo event generator. 
Subprograms used directly by the user are the following:
\begin{itemize}
\item 
  {\tt\bf KW\_ReaDataX}  
  -- the subprogram used to read, from the disk file, the default 
  input data of KoralW and subsequently the data of the user
  into the array {\tt xpar} at the very beginning of the use of KoralW.
  (This subprogram did not exist in the version 1.41.)
\item 
  {\tt\bf KW\_Initialize} 
  -- the subprogram that does all
  initializations of internal variables -- it calls several
  {\em initializers} of the main internal modules of the generator,
  such as {\tt karlud(-1,...)}, and of TAUOLA and 
  the external matrix element.
  It prints out directly or indirectly all the input parameters.
  (This subprogram replaces the {\tt koralw(-1,xpar,npar)} entry 
  in version 1.41.)
\item 
  {\tt\bf KW\_Make} 
  -- the most important subprogram of KoralW. It generates single MC events.
  Functionally it is the high-level management subprogram in the 
  event generation. 
  It invokes:
  \begin{enumerate}
  \item
    {\tt karlud},
    which provides the four-momenta of the outgoing fermions and 
    the ISR photons along with the value of the crude distribution,
  \item
    {\tt KW\_model\_4f}, 
    which computes the Born matrix elements either CC03 or CC/NC-all, 
    and all additional effects requested by the user, 
    e.g. anomalous couplings and/or the Coulomb correction,
  \item
    {\tt betar}, 
    which calculates the QED ISR corrections up to third order,
  \item
    {\tt tohep, tohad},
     the programs from TAUOLA, PHOTOS and JETSET libraries, which perform,
     respectively, $\tau$-decays, generation of bremsstrahlung in $W$ 
     and $\tau$ decays, and hadronization.
  \end{enumerate}
  {\tt KW\_Make} also decides about the rejection of an event
  in the case of a constant-weight event, or of a semi-constant-weight event.
\item 
  {\tt\bf KW\_Finalize} 
  does all final bookkeeping, including the calculation of the integrated 
  (total) cross section. It prints a summary output on all series of generated 
  MC events.
  (This subprogram replaces the {\tt koralw(1,xpar,npar)} entry 
  in version 1.41.)
\end{itemize}

The other important internal units of the generator are:
\begin{itemize}
\item 
  {\tt \bf karlud}
  The routine {\tt karlud} manages the actual phase-space point
  generation. 
  Upon initialization in {\tt mode}=$-$1, the event generation sequence in 
  {\tt mode}=0 
  is the following:
  \begin{enumerate}
  \item
    {\tt yfsgen} generates the $s'$ variable and the ISR photons in the LAB 
    frame.
  \item
    {\tt decay} generates the decay channel based on pretabulated
    cross-sections (and stores its ID number in the variable {\tt label}).
  \item
    {\tt make\_phsp\_point\_*} 
    generates four-vectors of fermions depending on
    the type of the final states drawn. It takes place in the effective CMS
    frame, i.e. the rest frame of the outgoing fermions.
    There are two different generators provided in this place.
  \item
    {\tt from\_cms\_eff} 
    transforms the four-momenta of the event from 
    the effective CMS to the LAB frame.
  \item
    {\tt selecto} 
    imposes cuts on the four-momenta. 
    The {\tt selecto} routine contains the
    built-in cut-offs, whereas the {\tt user\_selecto} routine is provided
    for the user-defined cuts -- it is located in the file 
    {\tt demo.14x/user\_selecto.f} and by default no cuts are applied there. 
    It is a matter of efficiency that as many
    of the cuts as possible should be imposed in this low-level routine to
    avoid further, time-consuming, steps of event generation for
    events that are to be rejected.  
  \item
    {\tt get\_phsp\_weight\_*} 
    calculates the crude distribution (the crude weight) for the 
    accepted event.
  \item
    {\tt karlud} 
    calculates necessary overall normalization factors.
  \end{enumerate}
  {\tt mode}= 1 and 2 provide, as usual, some post-generation information.

  In the following {\tt *} is a wildcard for a part of a routine name. 
\item 
  {\tt \bf make\_phsp\_point\_*}
  calls the {\tt *\_brancher} routines that do the random choice of 
  kinematical branch to be generated according to the preset probabilities. 
  It should be noted that
  these probabilities are dummy parameters and {\em can} be changed by
  the user, e.g. to cross-check the program or to fine-tune its efficiency. 
  It is also possible to add some more branches to the generator.
  Next, {\tt make\_phsp\_point\_*} calls the {\tt *\_spacegen} routines in
  order to build the actual four-momenta.
\item 
  {\tt \bf get\_phsp\_weight\_*}
  performs the summation over various branches of the multi-branch 
  four-fermion crude distribution in order to construct the total 
  normalization of the event. It also uses the {\tt *\_spacegen} routines.
\item 
  {\tt \bf *\_spacegen}. 
  The actual hard-core of the generation and normalization of the 
  four-fermion phase-space point.
  It is based on a complicated series of 
  kinematical-variables generation. 
  It is performed in various frames, in various
  orders and with different types of singular behaviour, depending on the final
  state and, subsequently, on the branch chosen at a given time by the
  program on previous stages of event generation. Details of the
  algorithm will be presented elsewhere \cite{jadach:preparation}. 
\end{itemize}

\subsection{{\tt B.E.} directory for the Bose-Einstein effect}

The package of programs in subdirectory {\tt B.E.}
implements the Bose-Einstein (BE) effect in the hadronization using,
the ``weight method'' described in Ref.~\cite{JadachZalewski:1997}.
It is not the integral part of the MC generator but rather a stand-alone
application of KoralW.
It has a double role: 
(1)~to illustrate how to implement the Bose-Einstein effect
according to the method of Ref.~\cite{JadachZalewski:1997},
(2) to provide an example (template) of the use of KoralW in a C++ environment.
The entire exercise on BE is implemented in C++.
KoralW provides ``raw'' events, which are translated to C++ structures, and the
whole process of constructing the BE weight, analysing events (i.e.
defining jets, eliminating combinatorial background, fitting the $W$ mass),
histogramming  and graphical output is done in C++.
There are only several lines of Fortran code (in {\tt ReaData.f}) 
in the entire
subdirectory {\tt B.E.}
The histogramming, fitting and graphics are done with the novel ROOT system
\cite{root:1998}, also entirely in C++.

The topography of the {\tt B.E.} directory is the following:\\
\hbox{\hspace{2cm}}
\begin{tabular}{ll}
{\tt B.E./src}  &--   sources and compilation/link objects,\\
{\tt B.E./run}  &--   data files and outputs for/from runs,\\
{\tt B.E./fig}  &--   analysing results from {\tt ../run}, 
                      all plotting/fitting, etc.,\\
{\tt B.E./bak}  &--   attics.\\
\end{tabular}\\
See also {\tt README} file in {\tt B.E.} directory.

The source files and executables are in {\tt B.E./src} directory.
Let us very briefly characterize their functionality:
\begin{itemize}
\item 
  {\tt rmain.C} is the main program. It runs the main loop over events.
\item 
  {\tt Semaph} class is used by {\tt rmain.C} to manage main loop over events
  using information from ``semaphore'' disk files%
  \footnote{The {\tt Semaph} class is a translation from the analogous F77 
  subprogram in {\tt BHLUMI} \cite{bhlumi4:1996}.}.
\item 
  {\tt ROBOL} class procedures are called in {\tt rmain.C}.
  It manages the generation of events, the calculation of the BE weight, 
  and the analysis of MC events, all by dedicated classes.
\item 
  {\tt KoralwMaker} class provides for the C++/F77 interface 
  with the {\tt KoralW} event generator.
  It transfers input data from the main program, which is in C++,
  to the {\tt KoralW} generator, which is in F77.
  It also picks up the output data (events) from {\tt KoralW} and makes 
  them accessible to {\tt ROBOL} and other C++ classes.
\item 
  {\tt KorEvent} is a class for the single MC event from {\tt KoralW}, 
  with some additional data fields/members for the analysis.
  The {\tt ROBOL} procedures handle the current MC event as a {\tt KorEvent} 
  object.
\item
  {\tt PartLund} is a class for a single particle in a format very close
  to the PDG/LUND convention --
  a single event of the type {\tt KorEvent} is simply a list
  of objects of the type {\tt PartLund}.
\item
  {\tt VLorentz} is a class for describing the single Lorentz four-momentum.
  It provides some basic functionality such as linear algebra and dot-product.
  It is used to construct the type {\tt PartLund} 
  in other places of the package.
\item
  {\tt BEwtMaker} is a class in which the current event defined by 
  the {\tt KoralwMaker} object gets assigned the BE weight.
\item
  {\tt JetAnalyzer} is a class in which the current event undergoes standard 
  analysis -- four jets are defined, the combinatorial background is 
  eliminated and the masses of jet pairs
  are recorded in histograms/{\it n}tuples.
\end{itemize}
Note that F77 externals are referred to {\em only} in {\tt KoralwMaker} 
and {\tt Semaph} classes.
The only extra F77 code is composed of two tiny routines in {\tt ReaData.f}
(they interface {\tt OPEN/CLOSE} functions of F77).

The input data for the run and the resulting histograms
are stored in {\tt run} subdirectory.
In particular the {\tt run/172GeV.4J} is the most important subdirectory.

The final analysis of the MC results is done
in {\tt fig} subdirectory, using several C++/CINT scripts
processed by ROOT.
In particular all eight figures of Ref.~\cite{JadachZalewski:1997}
are produced by them.
For instance the plot with the fit of the two-jet distribution 
with the Breit-Wigner formula
is made by the macro {\tt view-FitJetBE.C}.
It reads histograms from the {\tt run/200GeV.4J/rmain.root} file,
which can be produced with ``{\tt make 172GeV.4J-start}''.
For more details on
how to run the production-analysis sequence of the programs, see
the files  {\tt README} and {\tt src/Makefile}.
Let us finally advertise that
it is possible, in particular, to get a menu of graphics programs
with the command ``{\tt make fig}'' and to produce the on-line documentation
of all our C++ classes in the html format and view it with the html browser
with a single command: ``{\tt make dok-view}''.

In the preparation on the F77/C++ interface we profited a lot from inspecting
the C++ code of ATLFAST \cite{atlfast} and from discussions with the authors 
of this program.
Helpful discussions with the authors of the ROOT system are also acknowledged.

\section{Details of the Program}

\subsection{Four-Fermion Phase-Space Generation }

The generation of the four-fermion phase space in KoralW
is based on a multibranch type of Monte Carlo algorithm, 
cf.\ e.g.\ 
\cite{tauola:1992,tauola:1993,kleiss:1993,berends:1994,excalibur:1995}. The
cross-section is calculated in the usual way, as an average of the ratio of the
exact matrix element to the crude one, averaged over the crude distribution:
\begin{eqnarray}
\sigma &=& \int d{\rm Phsp}\;   \vert M \vert^2 
       = \left< {{ \vert M \vert^2}\over \tilde f_{CR}} \right>_{d\tilde\rho}
           \int d\tilde\rho, 
\\
d\tilde\rho &=&  d{\rm Phsp}\;   \tilde f_{CR}
 = \sum_i^{\rm branches} d{\rm Phsp}^i\; p_i\; \tilde f_{CR}^i ,
\label{DPS}
\end{eqnarray}
where $p_i$ is a probability of generating the branch $i$ and
$d{\rm Phsp}^i$ is already parametrized by angles and masses defined 
in branch-dependent Lorentz frames:
\begin{equation}
d{\rm Phsp}^i
   = ds_1^ids_2^i \prod_{j=1,2,3} d\cos\theta_j^i\; d\phi_j^i\; \lambda_j^i.
\end{equation}
Equation (\ref{DPS}) describes a general framework for generating arbitrary
phase-space configurations.  
The process specific information (and the difficulty!) 
is hidden in the $\tilde f^i_{CR}$ functions. Each branch 
of $\tilde f^i_{CR}$ is designed to describe a certain type of
singularity that is encountered in the Feynman graphs. 
As an example, the $t$-channel singularities can be thought of as $1/t$, 
as the $s$-channel ones as resonances, etc.
But this is of
course process-dependent. In the case of four-fermion final states
there is nearly a hundred different possible final states, 
each of them having up to
over a hundred Feynman graphs in the matrix element, each graph
having a different structure of singularities. Of
course there is a lot of symmetries and similarities that can be employed
in constructing the generator branches. Nonetheless, the number of different
branches exceeds fifty in the case of KoralW. In fact each branch
$i$ of Eq.~(\ref{DPS}) consists of many ``sub-branches'', which split on
subsequent lower
levels of generation. The total number of such ``elementary'' branches
would exceed a million. Just by looking at this large number, one can
realize that optimalization must play an essential role in the
algorithm. Indeed, there is a large number of internal parameters and
coefficients, as $p_i$ of Eq.~(\ref{DPS}), that must be
fine-tuned in order to enhance the desired singularities and damp the
others for given final-state configurations.

In the KoralW code we have implemented two different sets of
libraries of these $\tilde f^i_{CR}$ functions. 
In addition there is a third possibility of using both libraries
simultaneously in a stochastic mixture.
These packages provide a powerful tool for controlling the
complicated and sometimes numerically unstable integration. 
Moreover, the coefficients $p_i$ of Eq.~(\ref{DPS})
are dummy parameters of the generators and can be changed by
the user for the consistency tests.

Details of the actual algorithms will be presented elsewhere
\cite{jadach:preparation}.

\subsection{External Matrix Element}

In its present version, KoralW includes an interface to the external
library calculating the correction weight due to a different, external,  
matrix element. 
The idea behind this is that the user may occasionally 
wish to replace the internal matrix element by a different one, for
instance including special combinations of the anomalous couplings.
Thanks to the modular structure of KoralW and, in particular,
to the factorizability%
\footnote{
It is important to stress that this {\em practical} factorizability of
the algorithm is due to the approximation in the treatment of complete
bremsstrahlung. Omitted non-factorizable corrections may produce severe
effects in certain final-state configurations, see Section \ref{NONISR}
for details.
}
of the
approximate QED matrix element into the Born matrix element and the QED part,
it is straightforward
to replace the existing Born-level matrix element  with any other one,
provided that the external library is able to calculate the corresponding
matrix elements out of the
externally generated four-momenta. To this end the external program 
provided by the user must be able to calculate the matrix element,
completely normalized (in picobarns) but without any additional
factors such as flux-factor $1/(2s')$, etc.
In the following subsections we will explain 
how to use the interface and describe in more detail the external
libraries supplied with KoralW. 

\subsubsection{Interface}

A predefined interface, now included in KoralW,  will activate user-supplied 
routines with the help of the {\tt Key4f} key.
For {\tt Key4f=0} no 
external matrix element is
included and for {\tt Key4f=1 } it is active. The new position  of the weight
switch {\tt KeyWgt} is also introduced. 
For {\tt KeyWgt=2}, the generation is done in two steps: in the first step 
the program
generates constant-weight events according to CC03 distribution and 
in the second
step the external weights are calculated and transmitted
to the common block {\tt /wgtall/} without rejection, i.e.
the user gets variable-weight events.

In order to avoid possible overwriting errors in the execution, 
the communication
of the libraries with the main generator is done through dedicated 
interface routines. 
On the side of the generator the file {\tt amp4f\_ini.f} 
in the directory {\tt interfaces} contains the ``buffer'' subprograms.
On the user side, his/her own directory has to replace the directory 
{\tt ampli4f}.
The following two routines have to be provided by the user:
\begin{enumerate}
\item
  {\tt ampini(xpar,npar)},
  which initializes the external matrix element library. 
  The standard input parameter array {\tt xpar} can be used there for 
  initialization purposes (entries above 4000).
  The additional parameters are made available to the user 
  by calling the subroutine {\tt masow(sin2w,gpicb,amaf)}
  with {\tt sin2w} being the $\sin^2\theta_W$, {\tt gpicob} the conversion
  factor to picobarns and {\tt amaf(20)} the matrix of fermion masses.
\item
  {\tt amp4f(q1,ifbm1,q2,ifbm2,p1,ifl1,p2,ifl2,p3,ifl3,p4,ifl4,wtmd4f,wt4f)}\\
  should  calculate the new matrix element squared {\tt wtmd4f}, fully
  normalized, but without any additional factors such as e.g. 
  flux factor $1/(2s')$. The arguments \\
  {\tt q1,ifbm1,q2,ifbm2,p1,ifl1,p2,ifl2,p3,ifl3,p4,ifl4} denote
  respectively four-momenta and identifiers (according to the
  PDG conventions \cite{PDG:1990}) of the initial-state effective
  beams and the final-state fermions before the final-state bremsstrahlung
  generation. In fact the four-momenta of the final-state fermions are always
  supplied by KoralW to {\tt amp4f} in the same order as in the
  {\tt /cms\_eff\_momdec/} common block. 
  The additional vector of weights {\tt wt4f(i), i=1,9} may optionally
  be filled in by the routine
  {\tt  amp4f}. It is not used in the program but only
  transmitted to the KoralW optional weights common block
  {\tt /wgtall/} as {\tt wtset(40+i)}. The {\tt wtmd4f} is put into
  {\tt wtset(40)}.
\end{enumerate}

\subsubsection{All Four-Fermion Library of GRACE v. 2}
\label{4flib}

A working example of the above interfacing philosophy is the library of all
four-fermion matrix elements. It is produced by the package for
automated calculations GRACE version 2 \cite{GRACE2} and provided for KoralW
by courtesy of the Minami-Tateya Group of KEK.
It consists of a set of fully massive matrix elements for all possible 
four-fermion processes in the $e^+e^-$ collisions. 
Let us describe some features of the package:
\begin{enumerate}
\item
  The CKM matrix is diagonal. For the non-diagonal CC final states the
  internal CC03 massless matrix element is used as a temporary fix.
  For the non-diagonal, doubly-CKM-suppressed MIX-type final states, i.e.\
  simultaneously CC- and NC-type
  ($u\bar ss\bar u, u\bar bb\bar u, c\bar dd\bar c, c\bar bb\bar c$), 
  the external matrix element (it is without CC03 graphs due to the diagonal
  structure of the CKM matrix) 
  is incoherently added to the internal one
  (containing CC03 graphs only) and the interference of these
  two is neglected.
\item
  It can be downgraded to the CC03 case with the help of the dipswitch
  {\tt iswitch=0} in the {\tt amp4f} routine.
\item
  There are Higgs exchange-graphs included, 
  activated by the dipswitch {\tt jhiggs} in the file
  {\tt grc4f\_init/setmas\_koralw.f}, cf. GRACE manual
  \cite{GRACE2} for details of the model.
\item
  The gluon-exchange graphs can be activated with the help of the dipswitch
  {\tt jgluon} in the file
  {\tt grc4f\_init/setmas\_koralw.f}, cf. GRACE manual
  \cite{GRACE2} for details.
\end{enumerate}

\subsubsection{CC-All Library of GRACE v. 1}

\label{old-grace}
There is yet another external library distributed with KoralW. 
It is a library of massive matrix elements for the CC-type processes. 
It is based on the earlier version 1 of the GRACE package \cite{GRACE1}. The
important difference, as compared with the previous library, 
is that complete set of scripts can be 
found in here to generate the library from scratch,
i.e. to build it by the algebraic program GRACE and then custom-fit 
for the KoralW. It requires the GRACE v.~1 package to
be installed. Also, the complicated structure of makefiles and scripts
may pose some portability problems. We have developed the package on an
HP735 computer running  HP-UX v. 9.x.

Currently, this library is obviously inferior to the
GRACE v. 2 library, and we {\em do not} recommend using
it for the standard applications.
On the other hand, for an advanced user, it gives a possibility 
to define his/her own modified ``Standard Model'' at the
Lagrangian level (within the GRACE system) 
and then directly have a ready-to-plug-in code generated. 
For this and some other reasons we decided to keep it in the
distribution version.
We recommend the interested users to contact us directly for more information
on its use.

\subsubsection{Cuts, Instabilities and Precision}
\label{CUTS}
The following set of primary cuts at the low level of MC
generation is introduced in KoralW v. 1.42 (in subroutine {\tt selecto} 
in {\tt korww/karludw.f}):
\begin{enumerate}
\item
  on the invariant-mass squared of produced pairs in final states with 
  two or more electrons/positrons, 
  $e^+(p_1)e^-(p_2)f(p_3)\bar{f}(p_4)$:
  $(p_1+p_2)^2 > $ {\tt arbitr1},
  $(p_3+p_4)^2 > $ {\tt arbitr1};
  in the case of $ e^+(p_1)e^-(p_2)e^+(p_3)e^-(p_4)$ final state, 
  additionally, 
  $(p_1+p_4)^2 > $ {\tt arbitr1},
  $(p_2+p_3)^2 > $ {\tt arbitr1}
  are required;
\item
  on angles of charged fermions with respect to the beams:\\
  $\angle(p_i,\hbox{beam}_j)> $ {\tt themin} [rad]
\item
  on transverse momenta of visible charged fermions
  $p_T^2 >$ {\tt arbitr}
  with $p_T^2$ defined as
  $p_T^2=  \sum_i^4 \left(p_i^2(1)+p_i^2(2) \right)$ 
  summed over all charged fermions with the angles greater than {\tt
  themin} with respect to the beams;
\item
  on transverse momenta of photons for $e^+e^-f\bar{f}$-type final states:\\
  $\left(\sum_i^4 p_i(1)\right)^2 + 
  \left(\sum_i^4 p_i(2)\right)^2 < $ {\tt arbitr2}\\
\end{enumerate}
The default and recommended values of these cuts are:\\
{\tt  arbitr }= 600 GeV$^2$  for  minimal visible $p_T^2$, \\
{\tt arbitr1 }= 8   GeV$^2$  for  minimal invariant mass in $e^+e^-f\bar{f}$, \\
{\tt themin }=  1$\times 10^{-6}$ rad for  minimal angle with beam, \\
{\tt arbitr2 }=  300 GeV$^2$ for maximal $p_T^2$ of photons in $e^+e^-f\bar{f}$. \\
Let us now turn to the basic question: Do we need these cuts and why?
Before we try to answer it, let us stress that these cuts {\em
can} be removed completely. Both the presamplers cover the {\em entire}
four-fermion phase space, no part of it is cut-out during 
the generation process. 
The above cuts are imposed {\em after} the generation is
completed, but {\it before} the matrix element is calculated. Now, coming back
to the basic question, there are two
reasons for non-zero recommended values of the cuts:
\begin{itemize}
\item
  Technical problems in the evaluation of the external matrix element.
  It can be numerically unstable for some extreme four-momenta 
  configurations (when some of the invariants become very small). 
  \begin{enumerate}
  \item
    For the CC-type processes we have {\em not} encountered 
    any instabilities in any region of the phase space while
    using presampler No.~1 in the entire phase-space, so no cuts are
    needed%
    \footnote{
      Of course we cannot guarantee that these instabilities would not show up
      with some combination of input parameters, with sufficiently long series
      of generated events or even on some platforms other than the ones 
      we tried.
    }.
    In the case of presampler No.~2 we noticed some residual instabilities
    that can be cut out with the angular cut {\tt themin }=  1$\times 10^{-6}$
    rad. This cut influences the total cross-section at the level 
    of 1-2 per mille 
    and therefore can be neglected within current precision requirements.
    (The presence of instabilities only in presampler No.~2 is due to the fact 
    that this presampler is more oriented on the singular configurations and 
    in a sense reaches ``closer'' to the singularities than the other one.)
  \item
    The other two cuts are for the  processes classified as NC-type. 
    The instabilities
    show up either for small transfers in the $t$-channel-dominated
    configurations (cut out by {\tt arbitr}) or for small invariant masses of 
    fermion pairs (cut out by {\tt arbitr1}). Note that the {\tt arbitr}
    cut, although defined by transverse momenta, is intended to assure
    that the energy of the (other and undetected) final-state electron, even 
    if lost in the 
    beam pipe, is noticeably smaller than the beam energy.

  \item
    The above three cuts have proved to us to be sufficient to deal with
    the Born-level
    instabilities in the matrix element (see previous footnote). 
    Unfortunately the picture gets
    more complicated in the presence of ISR bremsstrahlung. Namely, with the
    emission of the transverse photons the effective CMS frame that we use to
    define the four-momenta for the matrix element calculation gets 
    rotated with respect to the LAB frame in which the cuts are imposed 
    (see Ref.~\cite{koralw:1995a} for details on the construction 
    of this effective frame). 
    As a result, a ``large-angle-event'' in the LAB frame can lead 
    to the highly collinear and numerically unstable matrix-element 
    calculation in the effective CMS frame. For that reason the 
    {\tt arbitr2} cut is introduced along with the relatively high values of 
    {\tt arbitr} and {\tt arbitr1} cuts.   
  \end{enumerate}
  In principle, the instabilities of points 1 and 2 can be cured with 
  the help of
  quadruple-precision arithmetics. It is an easy task on the IBM or SGI
  workstations as they support {\tt COMPLEX*32} type and the upgrade can
  be done easily with the help of the compiler options. 
  This cures the technical side of the problem. However,
  the problem described in point 3 will remain unsolved, as
  the cross-section (and weights!), although numerically 
  stable, will be huge and very likely physically wrong in the 
  presence of transverse photons. 
  It is due to the ill-defined simulation of the missing 
  complete ${\cal O}(\alpha)$ 
  matrix element and the use of the effective beams technique. 
\item
  Physical problems.
  At the end of the last point we already presented one aspect of physical
  inaccuracy due to the lack of a complete ${\cal O}(\alpha)$ matrix
  element. The other related fundamental problem is due to the fact that the
  bremsstrahlung process is treated in the Initial State Radiation 
  approximation. It means that only
  the $s$-channel type radiation is generated. In the case of
  electrons in the final state, one often encounters for example 
  the Bhabha-like
  configurations that are dominated by $t$-channel-type diagrams. The
  radiation, however, is neither generated according to nor corrected for 
  these configurations. 
  This can lead to severe inaccuracies in the results. We
  refer the reader to Section \ref{NONISR} for more information.
\end{itemize}

To summarize, the technical instabilities in the Born cross section can be
cured either by going to the quadruple precision or by implementing
relatively modest cuts, weaker than the recommended ones. 
The inclusion of bremsstrahlung makes the picture complicated.
Because of the limitations of the effective beams technique, even the 
recommended strong cuts do not remove the instabilities completely. 
On the other hand, even if 
the quadruple precision is used one is still limited, in loosening cuts,
by the physical meaningfulness of the results, and, on the technical side, 
by large overweights due to overestimated cross-sections.
This subject requires further study, see also Section~\ref{NONISR}.

Let us also note that in the case of the CC03-type presampler
all the cuts are reset by the program to no-cuts values.

Together with the predefined cuts in the {\tt selecto} routine, we provide
another routine, {\tt user\_selecto}, in which the user can implement
his/her own cuts. This routine is located in the same file {\tt
korww/karludw.f}. 
If the user requires some stringent additional (pre)cuts they can be 
placed there. 
This may increase the speed of the program (depending on how many events
are rejected by the cuts) as the matrix element 
will not be calculated for unaccepted events.  
The user interested in the complete or nearly complete phase-space coverage
should start with setting all cuts to zero and inspecting overweighted events. 
(The {\tt arbitr2}, if set to a negative or zero value, will be reset by 
the program to its maximal value, i.e. the value of the $s$ parameter). 
The maximal weights for rejection can be adjusted in the input parameters
later on.

As an example of such a routine, we provide the routine 
{\tt user\_selecto\_canonical} with the canonical cuts as defined at the
LEP2 Workshop \cite{yr:lep2}. With its help we have for example reproduced,
within an accuracy of a few per mille, the results given in Tables~6-8 
(see p.~241 of Vol.~1) in Ref.~\cite{yr:lep2} (with the exception
of entry 14 of Table~7).

\subsection{Bremsstrahlung}

In the present version of our program we upgraded the ISR exponentiated matrix
element with third-order LL corrections.
We know that these effects are small, especially for the 
Yennie-Frautschi-Suura-type
exponentiation we use, see Ref.~\cite{jadach:1991}; 
this new contribution may nevertheless be useful for estimating theoretical
systematics due to QED higher orders.
The reader should keep in mind that the new third-order LL corrections 
influence longitudinal momenta of ISR photons and add nothing new in the $p_T$ 
distributions of the ISR photons.
Let us write the master equation for the total cross section,
with new terms due to third-order LL corrections:
\begin{equation}
\label{master}
\begin{split}
  \sigma =&
  \sum_{n=0}^\infty {1\over n!}
  \lint \prod_{i=1}^4 {d^3q_i\over q_i^0} \;
  \left(\prod_{i=1}^n       {d^3k_i\over k^0_i}
    \tilde{S}(p_1,p_2,k_i) \right)
  \delta^{(4)}\bigg(p_1+p_2-\sum_{i=1}^4 q_i
  -\sum_{i=1}^n k_i \bigg)
  \Theta^{cm}_\epsilon\;\;
\\
  &
  \exp\bigg(  2\alpha \Re B
              + \lint {d^3k\over k^0} \tilde{S}(p_1,p_2,k)
              (1-\Theta^{cm}_\epsilon)
      \bigg)
  \Bigg[
         \bbeta^{(3)}_0(p^\Rcal_r,q^\Rcal_s)
        +\sum_{i=1}^n
        {\bbeta^{(3)}_{1}( p^\Rcal_r, q^\Rcal_s, k_i) \over \tilde{S}(k_i)}
\\&
        +\sum_{ i>j }^n
        {\bbeta^{(3)}_{2}( p^\Rcal_r, q^\Rcal_s, k_i, k_j)
                             \over \tilde{S}(k_i)\tilde{S}(k_j)}
        +\sum_{ i>j>l }^n
        {\bbeta^{(3)}_{3}( p^\Rcal_r, q^\Rcal_s, k_i, k_j, k_l)
                             \over \tilde{S}(k_i) \tilde{S}(k_j) \tilde{S}(k_l) }
  \Bigg],
\end{split}
\end{equation}
where $\tilde{S}(p_1,p_2,k) = -(\alpha/ 4\pi^2)
\big( (p_1/ kp_1) -(p_2/ kp_2) \big)^2$
is the real photon infrared (IR) factor and
$\Theta^{cm}_\epsilon=\prod\limits_{i=1}^n 
\theta\left(k^0_i-\epsilon\sqrt{s}/2\right)$ cuts out the singular IR
region, already included to all orders in the YFS form factor 
\cite{yfs:1961,koralw:1995a}.
In Eq.~(\ref{master}) we use essentially the same notation as in 
Ref.~\cite{koralw:1995a},
except for $p^\Rcal_r={\cal R}p_r,\; r=1,2$ 
and $q^\Rcal_s={\cal R}q_s,\;s=1,\,...,\,4$.
Note that the ``reduction procedure'' ${\cal R}$ does not act on photons.
The new third-order non-IR functions
$\bbeta^{(3)}_{l},\; l=0,\,...,\,3$ are defined below.
In our \Order{\alpha^3} LL calculation all $\bbeta$'s are proportional to
the basic lowest-order (Born) distribution
\begin{equation}
  b_0^\Rcal
  =b_0(p^\Rcal_i,q^\Rcal_j)=
  \bbeta^{(0)}_0(p^\Rcal_1, p^\Rcal_2, q^\Rcal_1,..., q^\Rcal_4 )
  = {1\over 2s'}\;
  {(2\pi)^4 \over \left[ 2(2\pi)^3 \right]^4}\;
  \sum_{\rm spin} |{\cal M}_{\rm Born}|^2,
\end{equation}
in which fermion momenta are adjusted (with $\Rcal$-procedure) in such 
a way that the 
Born matrix element ${\cal M}_{\rm Born}$ is calculated at the reduced
centre-of-mass energy $s'=(p_1+p_2-\sum_{i=1}^n k_i)^2 $.

\subsubsection{ISR up to Third Order}

The \Order{\alpha^r} functions $\bbeta_i^{(r)}$, see \cite{yfs:1961},
are residuals of the removal of IR virtual and real 
singularities. They are therefore IR-finite.
In the YFS \cite{yfs:1961} scheme, 
they are obtained from the \Order{\alpha^r}
``raw'' differential distributions, which originate from the Feynman diagrams.
This is also the case in the  \Order{\alpha^1} YFS-exponentiated
matrix element of BHLUMI \cite{bhlumi2:1992}
or YFS2 \cite{yfs2:1990} and YFS3  \cite{yfs3:1992}.
Sometimes the more ``economical'' source of 
raw differential distributions is not the Feynman rules
but the LL approximation, especially for
\Order{\gamma^2} and \Order{\gamma^3} corrections, where
\begin{equation}
  \gamma = 2{\alpha\over\pi} \left( \ln{s\over m^2_{e}}-1\right).
\end{equation}
Here, in our calculation, we employ the LL approximation up to 
\Order{\gamma^3}.

The first step in our procedure of deducing the third-order LL exclusive
distribution is to examine
triple convolution of the Altarelli-Parisi (AP) non-singlet evolution equation.
This is done separately for two $e^\pm$ incoming lines and the
convolution of the
two results is examined once again (truncating the final result up to 
\Order{\gamma^3}).
It can also be done in one step, using a triple convolution
of the AP equation, with a ``double'' AP kernel acting on both  $e^\pm$ lines.
All the above is a one-dimensional exercise, i.e. variables in the AP equation
are interpreted as ratios of the photon energies to the $e^\pm$ energies, and
the photons are assumed to have exactly zero transverse momentum.
The above calculation is rather straightforward, quite similar to that
done in Ref.~\cite{skrzypek:1992}, and we shall here skip this part.
With the explicit longitudinal momentum of up to 3 photons,
we supplement them with the smooth $p_T$ distribution, getting in this way
what we call the ``LL ansatz'' for full differential distributions, 
with up to three real photons.
In the ansatz we are careful to reproduce all correct
soft-photon limits, 
i.e. for any configuration with any number of photons being soft.
In the above construction, virtual up to \Order{\gamma^3} corrections are kept
all the way through the construction procedure,
and the cancellation of IR singularities is kept perfect.
The resulting \Order{\gamma^3}-finite $p_T$ ansatz for $n_\gamma=0,1,2,3$
photons reads as follows:
\begin{equation}
\begin{split}
  & D_{[0]}^{(3)} = 
  \left\{  
      1+2B(1) +{1\over 2}[2B(1)]^2 +{1\over 6}[2B(1)]^3 
  \right\} b_0^\Rcal,\\
  & D_{[1]}^{(3)}(k_1)=
  \tilde{S}(k_1) \chi(\alpha_1,\beta_1)
  \bigg\{
      1+{1\over 2}[3B(1)+B(z_1)]^2 
\\&\qquad\qquad\qquad
       +{1\over 6}\left[ (2B(1))^2 +(2B(1))(B(1)+B(z_1)) + (2B(z_1))^2 \right]
  \bigg\}\; b_0^\Rcal,\\
  & D_{[2]}^{(3)}(k_1,k_2)=
  \tilde{S}(k_1)\tilde{S}(k_2) 
  \chi(\alpha_1,\beta_1) \chi(\alpha_2^*,\beta_2^*)
  \left\{
      {1\over 2} +{1\over 6}[4B(1) +B(z_1)+B(z_1z_2)]
  \right\}\; b_0^\Rcal,\\
  & D_{[3]}^{(3)}(k_1,k_2,k_3)=
  \tilde{S}(k_1)\tilde{S}(k_2)\tilde{S}(k_3)\; 
  \chi(\alpha_1,\beta_1) \chi(\alpha_2^*,\beta_2^*)\chi(\alpha_3^*,\beta_3^*)\; b_0^\Rcal,\\
  & \chi(a,b) \equiv ((1-a)^2 + (1-b)^2)/2,\\
  & B(z)= {\gamma\over 2} A(z),\quad A(z)= {3\over 4} +\ln\varepsilon - \ln z,
\end{split}
\end{equation}
where $\alpha_i$ and $\beta_i$ are the familiar Sudakov variables
\begin{equation}
  \alpha_i = (k_i p_1)/(p_1 p_2),\quad \beta_i = (k_i p_2)/(p_1 p_2)
\end{equation}
while ``starred'' variables take into account the effects 
of energy loss due to prior emissions
\begin{equation}
  \begin{split}
   &\alpha_2^* =  \alpha_2/(1-\alpha_1),\quad 
    \beta_2^*  =  \beta_2 /(1-\beta_1),\\
   &\alpha_3^* =  \alpha_3/(1-\alpha_1-\alpha_2),\quad 
    \beta_3^*  =  \beta_3 /(1-\beta_1-\beta_2),
  \end{split}
\end{equation}
and we define $z$-variables as
\begin{equation}
  z_1     = (1-\alpha_1)(1-\beta_1),\qquad
  z_1 z_2 = (1-\alpha_1-\alpha_2)(1-\beta_1-\beta_2).
\end{equation}
The $\varepsilon$ variable is an infrared cut-off (an IR-regulator in the 
AP kernel), which defines the lower limit of photon energy in our phase space, 
$k^0> \varepsilon \sqrt{s}/2$.
Note also that the fact that $\gamma$ comes from the angular integration 
over $\tilde{S}(k)$ is essential for the exact IR cancellations 
between the virtual and real photons.
The first step on the way to $\bbeta_{[n]}^{(3)}$ is to remove
IR virtual corrections (because they are already present 
in the YFS form factor)
\begin{equation}
  \beta_{[n]}^{(3)} =  
    \exp\left(-\gamma\left[ {1\over 4} +\ln\varepsilon \right] \right)
    D_{[n]}^{(3)}(k_1,...,k_n)\bigg|_{  {\cal O}(\gamma^3)  }
\end{equation}
and the resulting distributions (programmed this way in our code) 
read as follows:
\begin{equation}
\begin{split}
  & \beta_{[0]}^{(3)} = 
      \left\{1+{\gamma\over 2}
             +{1\over 2}\left[{\gamma\over 2} \right]^2 
             +{1\over 6}\left[{\gamma\over 2} \right]^3
      \right\}\; b_0^\Rcal,\\
  & \beta_{[1]}^{(3)}(k_1)=
  \tilde{S}(k_1) \chi(\alpha_1,\beta_1)
  \left\{
      1+{\gamma\over 2} +{1\over 2}\left[{\gamma\over 2} \right]^2
       -{\gamma\over 4} \ln z
       -{\gamma^2\over 8} \ln z +{\gamma^2\over 24} \ln^2 z 
  \right\}\; b_0^\Rcal,\\
  & \beta_{[2]}^{(3)}(k_1,k_2)=
  \tilde{S}(k_1)\tilde{S}(k_2) 
  \chi(\alpha_1,\beta_1) \chi(\alpha_2^*,\beta_2^*)
  \left\{
          1 +{\gamma\over 2} -{\gamma\over 6} \ln z_1  -{\gamma\over 6} 
                                                          \ln (z_1z_2) 
  \right\}\; b_0^\Rcal,\\
  & \beta_{[3]}^{(3)}(k_1,k_2,k_3)=
  \tilde{S}(k_1)\tilde{S}(k_2)\tilde{S}(k_3)\; 
  \chi(\alpha_1,\beta_1) \chi(\alpha_2^*,\beta_2^*)\chi(\alpha_3^*,\beta_3^*)
  \; b_0^\Rcal.
\end{split}
\end{equation}
We assume tacitly that Bose symmetrization is done in the above expressions, 
for $n=2,3$.
Finally, the real IR-parts are subtracted properly as follows:
\begin{equation}
\begin{split}
  & \bbeta_{0}^{(3)} = \beta_{[0]}^{(3)},\\
  & \bbeta_{1}^{(3)}(k_1)= \beta_{[1]}^{(3)}(k_1) 
                         -\tilde{S}(k_1) \bbeta_{0}^{(2)},\\
  & \bbeta_{2}^{(3)}(k_1,k_2)
      = \beta_{[2]}^{(3)}(k_1,k_2) 
        -\tilde{S}(k_1) \bbeta_{1}^{(2)}(k_2) 
        -\bbeta_{1}^{(2)}(k_1) \tilde{S}(k_2),
        -\tilde{S}(k_1)\tilde{S}(k_2)\bbeta_{0}^{(1)} \\
  & \bbeta_{3}^{(3)}(k_1,k_2,k_3)
      = \beta_{[3]}^{(3)}(k_1,k_2,k_3) 
\\&\qquad\qquad
       -\bbeta_{2}^{(2)}(k_1,k_2) \tilde{S}(k_3)  
       -\bbeta_{2}^{(2)}(k_1,k_3) \tilde{S}(k_2)  
       -\tilde{S}(k_1) \bbeta_{2}^{(2)}(k_2,k_3) 
\\&\qquad\qquad  
       -\bbeta_{1}^{(1)}(k_1) \tilde{S}(k_2) \tilde{S}(k_3) 
       -\tilde{S}(k_1) \bbeta_{1}^{(1)}(k_2) \tilde{S}(k_3) 
       -\tilde{S}(k_1) \tilde{S}(k_2) \bbeta_{1}^{(1)}(k_3)
\\&\qquad\qquad
       -\tilde{S}(k_1)\tilde{S}(k_2)\tilde{S}(k_3)\bbeta_{0}^{(0)}.
\end{split}
\end{equation}
The above subtraction is done numerically in the program.
This completes the definition of our \Order{\gamma^3} matrix element.
Let us note finally that the above parametrization was essentially copied from
the $\Keu\Keu$ Monte Carlo for fermion-pair production \cite{KK-MC}.

\subsubsection{Non-ISR Corrections}
\label{NONISR}
As was already mentioned in Subsection~\ref{CUTS}, in the processes with
at least one $e^+e^-$ pair in the final state some problems are encountered
when we try to include the leading QED corrections by 
convoluting the $s$-channel-type ISR bremsstrahlung with the Born-like
$e^+e^-\rightarrow e^+e^-f\bar{f}$ process, calculated 
in the ``effective beams''
rest frame, i.e. in the rest frame of the incoming $e^+e^-$, with reduced
energies after emission of the ISR photons. Generally, such an approach
leads to two kinds of problems\footnote{
                               Here, we do not discuss the problem 
                               that arises when any
                               of the electrons/positrons are lost in the
                               beam pipe; this requires special treatment.}: 
(1) too much radiation is generated, 
particularly in the high photon-$p_T$ range, and 
(2) the event weights are submitted to huge fluctuations. 
All this is because the cross section in the above 
4-fermion production channels is dominated by the low-angle Bhabha-like 
processes, where the $t$-channel $\gamma$ exchange plays a crucial role.
As a result, the photon radiation is governed by $\ln(|t|/m_e^2)$
rather than by $\ln(s/m_e^2)$. These two ``big logs'' differ by a factor
$\ln(s/|t|)$, which can be sizeable at low angles (in fact, it can itself be
regarded as a ``big log''); this factor can explain
an excess of the ISR bremsstrahlung when generated as in the $s$-channel.
The reason for too many high-$p_T$ ISR photons is that in the 
low-angle Bhabha-like process the interferences between the electron
and positron lines become very small,
while the destructive interference between initial- and final-state emissions
becomes strong, see e.g. Ref~\cite{bhlumi2:1992}.
As a result, the emission of high-$p_T$ photons is strongly suppressed,
while in the current implementation of the
$s$-channel-type of the ISR photon generation this interference is absent
and cannot damp the $p_T$ of photons.
The ultimate solution to this problem is the inclusion of 
the bremsstrahlung generation of both
the $s$-channel and $t$-channel in the
MC algorithm. This, however, is not easy to implement into the
current structure of the program\footnote{We consider this solution
                                 for the future versions of KoralW.}.  

The second kind of problems, i.e. the strong fluctuations of the event weights,
are partly connected to the first one. Namely, the emission of the
high-$p_T$ ISR photons can boost the fermions to very small angles
configuration, where the ``effective'' Born-like cross section 
is huge, even if, in the LAB frame, the final electrons/positrons
are emitted at moderate angles. This can also cause 
some numerical instabilities in the matrix element calculation,
as has been discussed in Subsection~\ref{CUTS}.   
This deficiency can be partly cured by solving the first problem, 
i.e. by damping the high-$p_T$ photon radiation. 
But even after this is done, some huge weight can
still appear as some residual high-$p_T$ photons still remain, 
e.g. in the configurations where only one of the electrons/positrons
is emitted at low angles. 

We could try to overcome the above difficulties by imposing
some additional cuts on the $e^+e^-f\bar{f}$ events, as described in  
Subsection~\ref{CUTS}. Our experience shows, however, that
even the presence of strong cuts does not solve the problems
completely, although it can stabilize the event generation
considerably. Besides, some of those cuts may not be easy to
implement in the real experiment.

In the mean time we are preparing an intermediate
solution, without going into the $t$-channel bremsstrahlung 
generation and/or the exact (off-shell) ${\cal O}(\alpha)$ matrix element,
which eliminates extra cuts.
We are constructing an additional weight
that would emulate the main features of the $t$-channel bremsstrahlung.
As a guideline, we use the YFS
approach \cite{yfs:1961} to the radiative Bhabha-like process,
looking at the behaviour of the hard bremsstrahlung matrix element 
at low angles.
Since this solution is not yet ready, we do not include it
in the current version of KoralW. Instead, we add a dummy
routine {\tt eexx\_wt\_cor} (the file {\tt model/eexxcor.f})
as an ``open slot'' where the real one will be ``plugged in'' when completed.

\subsection{Additional Corrections (CC03-Based)}

In this section we will describe the additional corrections that are
imposed on top of the Born-level $WW$ matrix element. All these
corrections have one common feature -- they are justified/calculated
for the CC03 subset of graphs only. However, it is of great importance to
have them incorporated, all at the same time, 
into the complete charged current (CC-all) matrix element in one way
or another. 
A detailed study on this subject 
has been presented in Ref.~\cite{koralw:1997} on the example of the 
KoralW and {\tt grc4f} codes. On the one hand, it proved that the
``common-sense interpolation'' does make sense, provided one performs
some minimum set of cross-checks. On the other hand, it showed
that indeed it is a ``code-dependent'' procedure and the
differences can be of the order of a few per mille. 
The possible choices for adding corrections are for example: 
additive versus multiplicative method or the amplitude level versus
the squared amplitude level.
In the following we will describe some of the effects: naive QCD, 
the non-diagonal CKM matrix, the Coulomb correction and the anomalous 
couplings.
In the future we hope that this list will be extended by the complete
${\cal O}(\alpha)$ corrections for the on-shell $WW$ production, 
available in the literature 
\cite{fleischer:1989,kolodziej:1991,bohm:1988,beenakker:1991,beenakker:1991a}
and already implemented by us in another Monte Carlo code -- YFSWW
\cite{yfsww2:1996,yfsww3:1998}.

\subsubsection{Naive QCD and CKM Matrix}

For the $WW$-type final states the naive QCD correction (NQCD) 
and the non-diagonal CKM matrix are 
 applied as a multiplicative
correction to the whole matrix element. In principle these corrections
are justified for the CC03 subset of graphs only. However, as the
 background graphs contribute very little at LEP energies, we follow the
standard approach \cite{yr:lep2} 
and apply them to the entire CC-all matrix element
\begin{equation}
\vert M_{CCall}(i,j) \vert^2 =
\vert M^{external}_{CCall}(i,j) \vert^2\;
\Reu^{CKM}_{NQCD}(i)\;
\Reu^{CKM}_{NQCD}(j)
\label{MATRIX_EL_1}
\end{equation}
where $\vert M_{CCall}(i,j) \vert^2 $ is the matrix element for the
$(i,j)$ decay channel of the $W$-pair as used by KoralW for
calculating weights and 
$\vert M^{external}_{CCall}(i,j) \vert^2 $ 
the external matrix element supplied to KoralW by the external library
for the $(i,j)$ decay channel of the $W$-pair.
We note in passing that in the CC03 case, the CKM and NQCD corrections
are already included in the  $\vert M_{CC03}(i,j) \vert^2 $ and there is
no need for the correcting factor $\Reu^{CKM}_{NQCD}$ in this case.
The correction
itself is calculated for each $W$ as the ratio of appropriate $W$ 
branching ratios:
\begin{equation}
\Reu^{CKM}_{NQCD}(i)=
{Br(i)\over Br(e\nu_e)} {Br_0(e\nu_e)\over Br_0(i)}, 
\end{equation}
with $Br_0$ being the ``bare'' $W$ branching ratios (1/3 and 1/9).
The physical $W$ branching ratios are set either directly to fixed
values ({\tt KeyBra=1}), corresponding to $\alpha_S=0.12$,
or calculated according to the formula of \cite{denner:1993}
 based on the CKM matrix ({\tt KeyBra=2}):
\begin{equation}
Br(q)= {1\over3} |V_{CKM}(q)|^2
{{1+{\alpha_S\over\pi}} \over {1+{2\over3} {\alpha_S\over\pi}}},
\quad
Br(l)= {1\over9}  
{1   \over {1+{2\over3} {\alpha_S\over\pi}}}.
\end{equation}
In either case the $\Gamma_W$ is forced to be recalculated by the
 program. The third option  {\tt KeyBra=0} is equivalent to {\tt
 KeyBra=2} with the diagonal CKM matrix (1/3 for quarks and 1/9 for leptons) 
and $\Gamma_W$ not recalculated.
The formula used for $\Gamma_W$ recalculation (default if the
 input value of $\Gamma_W$ is negative) is the following:
\begin{equation}
\Gamma_W = {{3G_\mu M_W^3} \over {2\sqrt{2} \pi}} 
\left( 1+{2\over3} {\alpha_S\over\pi} \right).
\end{equation}

Note that, unlike the situation in the previous versions of KoralW, 
here, the $W$
branching ratios are used solely for the normalization of the matrix
element.

For the $ZZ$-type final states, NQCD is applied directly as
a multiplicative correction 
\begin{equation}
\vert M_{NCall}(i,j) \vert^2 =
\vert M^{external}_{NCall}(i,j) \vert^2 
\Reu_{NQCD}(i)
\Reu_{NQCD}(j),
\end{equation}
\begin{equation}
\Reu_{NQCD}(q)= 1+ {\alpha_S\over \pi},\;\; 
\Reu_{NQCD}(l)= 1,
\end{equation}
to the whole matrix
element, for each generated quark pair. 
The $Z$-width is however {\it not} corrected automatically, 
and the user must take care of it by himself/herself (since it enters 
the program through the input parameters, it can easily be
set to the appropriate value).

\subsubsection{Coulomb Correction}

The Coulomb correction is taken from Ref.~\cite{khoze:1995}, Eq.~(9).
It is the first-order formula:
\begin{equation}
\Reu_{Coul}=1+{\alpha_{QED}\sqrt{s}\over 4p}
\left( \pi- 2\arctan \left( {\vert\kappa\vert^2 -p^2\over 2p\Re\kappa}
                     \right)
\right),
\end{equation}
\begin{equation}
p^2 = {1\over{4s}} ( s^2 -2s(s_1+s_2) +(s_1-s_2)^2 ),\;\;
E = {{s-4M_W^2}\over{4M_W}},
\end{equation}
\begin{equation}
\kappa =   \sqrt{{1\over2}M_W\left(\sqrt{E^2+\Gamma_W^2}-E\right)}  
        -i \sqrt{{1\over2}M_W\left(\sqrt{E^2+\Gamma_W^2}+E\right)}.
\end{equation}
In the case of the CC03 matrix element, $\Reu_{Coul}$ is a simple
over-all multiplicative correction.
In the case of the CC-all matrix element, it is implemented in 
the code in the form of an {\em additive} correction based on the CC03 matrix
element only, and Eq.~(\ref{MATRIX_EL_1}) now becomes:
\begin{equation}
 \vert M_{CCall}(i,j)  \vert^2 =
 \vert M^{external}_{CCall}(i,j)  \vert^2 
\Reu^{CKM}_{NQCD}(i)
\Reu^{CKM}_{NQCD}(j)
+ 
 \vert M_{CC03}(i,j)  \vert^2 (\Reu_{Coul} -1).
\label{MATRIX_EL_2}
\end{equation}
An identical correction is applied to the CC03 semi-analytical formula.

\subsubsection{Anomalous Couplings}

In the case of the complete charged current,
CC-all, matrix element, the anomalous couplings are 
implemented in the code in the form of an {\em additive} correction. Equations
(\ref{MATRIX_EL_1}) and (\ref{MATRIX_EL_2}) take on their final form:
\begin{eqnarray}
\vert M_{CCall}(i,j) \vert^2 =&
\vert M^{external}_{CCall}(i,j) \vert^2 
\Reu^{CKM}_{NQCD}(i)
\Reu^{CKM}_{NQCD}(j)
\nonumber \\
+&
\left(
\vert M^{ACC}_{CC03}(i,j) \vert^2 \Reu_{Coul} 
 -\vert M_{CC03}(i,j)  \vert^2
\right),
\label{MATRIX_EL_3}
\end{eqnarray}
which gives the complete description of the matrix element
of KoralW, including all the ``additional'' effects.

In the current version of the program we implement three
parametrizations of the anomalous $WWV$ couplings ($V = \gamma$ or $Z$). 
A particular parametrization  can be chosen by the user with the help of 
the input parameter switch {\tt KeyAcc}. Values of these couplings are
transferred to the program through the input parameters vector {\tt xpar},
and they have to be set up appropriately by the user in the initialization 
mode.  

The parametrizations of the triple gauge boson couplings are the following:
\begin{itemize}
\item
  {\tt KeyAcc=0}: the Standard Model (SM) (non-anomalous) couplings.
\item
  {\tt KeyAcc=1}:
  $\{g_1^V, \kappa_V, \lambda_V, g_4^V, g_5^V, \tilde{\kappa}_V, 
  \tilde{\lambda}_V\} \in {\mathbb C},\; (V = \gamma, Z)$
  \\
  The above set represents the most general parametrization of the $WWV$
  couplings that can be observable in the process where the vector
  bosons couple to effectively massless fermions \cite{hagiwara:1987};
  see also Ref.~\cite{TGC-YR:1996} and references therein.
  In general, there are 7 complex-number-valued couplings for each 
  $\gamma$ and $Z$, so altogether one needs to supply 14 complex 
  (or 28 real) numbers in this case.
  This can be done through the {\tt xpar} vector entries: 
  {\tt xpar(21-57)}, see Table~\ref{tab:KoralW-input2} for more details.
  The SM values for these couplings are:
  \begin{eqnarray}
    & & 
    g_1^V = \kappa_V = 1, 
    \nonumber \\
    & & 
    \lambda_V = g_4^V = g_5^V = \tilde{\kappa}_V = \tilde{\lambda}_V = 0. 
    \nonumber 
  \end{eqnarray}
  This parametrization has been present in KoralW since the version 1.03
  and is backward compatible. The new parametrizations added in this
  version of the program, for {\tt KeyAcc=2,3}, are described below.
\item
  {\tt KeyAcc=2}:
  $\{\delta_Z, x_{\gamma}, x_Z, y_{\gamma}, y_Z\} \in {\mathbb R}$\\
  This set of 5 real-number parameters represents deviations of 
  the $C$- and $P$-conserving couplings from their SM values;
  see Ref.~\cite{TGC-YR:1996} and references therein. 
  It can be related to the previous parametrization as follows:
  \begin{eqnarray}
    & & 
    g_1^{\gamma} = 1, 
    \nonumber \\
    & &  
    g_1^Z = 1 + \tan\theta_W\,\delta_Z, 
    \nonumber \\
    & &
    \kappa_{\gamma} = 1 + x_{\gamma}, 
    \nonumber \\
    & & 
    \kappa_Z = 1 + \tan\theta_W (x_Z + \delta_Z),   
    \nonumber \\
    & &
    \lambda_{\gamma} = y_{\gamma}, 
    \nonumber \\
    & &
    \lambda_Z = y_Z, 
    \nonumber \\
    & &
    \lambda_V = g_4^V = g_5^V = \tilde{\kappa}_V = \tilde{\lambda}_V = 0. 
    \nonumber
  \end{eqnarray}
  The values of the above 5 parameters are sent to the program through
  the {\tt xpar} vector entries: {\tt xpar(61-65)};
  see Table~\ref{tab:KoralW-input2} for details. 
  For the SM they are all {\it zero}.
  This parametrization is convenient for studies of the $C$- and 
  $P$-conserving $WWV$ couplings.  
\item
  {\tt KeyAcc=3}:
  $\{\alpha_{W\phi}, \alpha_{B\phi}, \alpha_{W}\} \in {\mathbb R}$\\
  This parametrization of 3 real-number-valued couplings
  corresponds to a direct extension of the SM formalism in terms
  of a {\em linear} realization of the symmetry, which can be
  achieved if a relatively light Higgs boson is assumed to exist,
  see Ref.~\cite{TGC-YR:1996} and references therein.  
  Its relation to the most general parametrization (for {\tt KeyAcc=1})
  reads:
  \begin{eqnarray}
    & & 
    g_1^{\gamma} = 1, 
    \nonumber \\
    & &  
    g_1^Z = 1 + \frac{\alpha_{W\phi}}{c_W^2}, 
    \nonumber \\
    & &
    \kappa_{\gamma} = 1 + \alpha_{W\phi} + \alpha_{B\phi}, 
    \nonumber \\
    & & 
    \kappa_Z = 1 + \alpha_{W\phi} - \frac{s_W^2}{c_W^2}\alpha_{B\phi},   
    \nonumber \\
    & &
    \lambda_{\gamma} = \lambda_Z = \alpha_{W}, 
    \nonumber \\
    & &
    \lambda_V = g_4^V = g_5^V = \tilde{\kappa}_V = \tilde{\lambda}_V = 0,
    \nonumber
  \end{eqnarray}
  where $s_W = \sin\theta_W,\,c_W = \cos\theta_W$. \\
  Values of the above 3 parameters are sent to the program through
  the {\tt xpar} vector entries: {\tt xpar(71-73)},
  see Table~\ref{tab:KoralW-input2} for details. For the SM they all are {\it zero}. 
\end{itemize}

\subsection{Renormalization Schemes}

The choice of renormalization scheme is a direct result of
the freedom in the renormalization procedure of the electroweak
sector. One can choose a different set of input parameters used to
express the cross-section. 
There is a large number of different schemes available in the literature;
see, for instance \cite{hollik:1993}.
All these schemes are more or less educated, but {\em ad hoc}
inclusions of some EW radiative corrections, and  are done by hand 
on the ``bare Born'' by means of improving its parameters. 
Therefore they can in principle vary from the ``bare'' (not corrected)
result and, among themselves, as much as 15\%; 
see for example \cite{fleischer:1989}.

In KoralW we have implemented three different schemes. Any of them
can be chosen with the help of the input parameter {\tt KeyMix}:
\begin{itemize}
\item
  {\tt KeyMix=0}
  ``LEP2 Workshop scheme'':\\
  $\alpha_W=$(input),\;
  $\sin^2\theta_W = \pi\alpha_W/(\sqrt{2} G_\mu M_W^2 )$,\;
  $g^2=4\pi\alpha_W/\sin^2\theta_W$;
\item
  {\tt KeyMix=1}
  ``G$_\mu$ scheme'':\\
  $\alpha_W=\sqrt{2} G_\mu M_W^2 \sin^2\theta_W /\pi$,\;
  $\sin^2\theta_W = 1-M_W^2/M_Z^2$,\;
  $g^2=4\pi\alpha_W/\sin^2\theta_W$;
\item
  {\tt KeyMix=2}
  ``bare Born'' (for tests only):\\
  $\alpha_W=\alpha_{QED}=1/137$,\;
  $\sin^2\theta_W = 1-M_W^2/M_Z^2$,\;
  $g^2=4\pi\alpha_W/\sin^2\theta_W$.
\end{itemize}

Note that this key has changed its meaning with respect to the earlier
versions of the KoralW code, where it simply changed the
definition of $\sin^2\theta_W$, leaving changes of the other input
parameters directly to the user.
Now it does the complete redefinition of the required 
input parameters according to the chosen scheme. 
The reader may have noticed that the options 
{\tt KeyMix=1} and {\tt KeyMix=2} have changed since the previous version,
whereas the setting {\tt KeyMix=0} remained unchanged.  

Finally, let us comment on the issue of the recommended setting.
The ultimate criterion of choosing the scheme 
would be the comparison with the exact electroweak corrections. 
This has been studied only for the on-shell case \cite{yr:lep2},
and no unique answer was given.
Favoured were, however, the ``Workshop'' or ``G$_\mu$'' type schemes.  
In fact, it would most likely be impossible to 
distinguish them experimentally,
as for example these two favoured schemes differ at the per mille level, far
less than the correction itself.  

\subsection{Colour (Re)Connection}

\subsubsection{Colour Connection}

The final state of KoralW generation consists of four fermions and
an arbitrary number of real photons. Later, external libraries are called,
if necessary, to perform, for instance, 
the $\tau$ decay and/or hadronization of the quark pairs. 
In most cases, the information stored by KoralW in the standard event record
{\tt COMMON /HEPEVT/} \cite{PDG:1990} is sufficient.

However, there exists a sub-class of four-quark final states
where the intermediate state can 
consist of two distinct configurations of colourless quark-pairs,
e.g. for the $u \bar u, d \bar d$ final state 
the colourless pairs of quarks chosen for hadronization
can be either $u \bar d$ and $d \bar u$ or $u \bar u$ and $d \bar d$.
This ambiguity has to be determined by KoralW, so that appropriate information
can be passed to the  hadronization package 
JETSET \cite{jetset:1987}. 

In our program, the random choice is made 
with the help of the weight
$$w_{CC} = { |M_1|^2 \over |M_1|^2 + |M_2|^2} \,,$$ 
where $M_1$ and $M_2$ 
represent spin amplitudes squared (and appropriately summed over 
spin degrees of freedom) for the two possible colour-singlet
configurations. The choice is performed with the help of a call to the routine 
{\tt  spdetx(ireco)}. The output parameter {\tt ireco=0,1} denote 
that the colourless object should be formed either from the
first-second and third-fourth quark pairs in the final state 
(as located in the {\tt COMMON /HEPEVT/}) 
or from the first-fourth and second-third quark pairs, respectively.

As an input for the routine {\tt  spdetx(ireco)}, the appropriate spin
amplitudes hidden in internal common blocks of the GRACE routines are used.
The routine {\tt  spdetx(ireco)} consists of the 
routine {\tt  spdetc} generated by the GRACE package \cite{GRACE2},
which is adapted to our purposes.

\subsubsection{Colour Reconnection}

Since in the LEP2 energy range the average distance between the $W^+$ 
and $W^-$ decay vertices 
is much smaller than the typical hadronic
size, the fragmentation of two $W$'s may not be independent.
One of the physical effects related to this problem is the colour
reconnection between decay products of different $W$'s\footnote{
          In the general case, not restricted just to the
          $W$-pair production and decay at the CC03 approximation, 
          the quark pairs, as explained above, can originate also
          from decays of $Z$'s or virtual $\gamma$'s, and the 
          quark-pair colour-singlet ambiguity can already exist in that step.
          Independently, in the second step, the colour arrangement 
          may need to be redefined, i.e. reconnected,
          owing to the final-state gluonic/hadronic interactions.}.
Several models have been proposed to describe this phenomenon
and estimate its influence on the $W$-mass measurement; 
see e.g. Ref.~\cite{Wmass-YR:1996} and references therein. 
The current version of KoralW does not include any of these models; 
it uses instead the user-supplied value of the colour-reconnection
probability {\tt PReco} (see Table~\ref{tab:KoralW-input1}) 
to generate quark-antiquark colour objects 
from the decay products of two different $W$'s. 
These objects are then processed by the JETSET routines to result 
in the final-state hadrons. 

This, rather simplistic, approach to the colour reconnection problem
does not deal, of course, with all the aspects of this phenomenon,
but it can be used for some simple estimate of its influence on physical 
observables.
In detailed studies, however, this has to be confronted with the results
of dedicated models.

\subsection{Bose-Einstein Effect in Hadronization}

It is generally expected that in the double hadronic decay of the $W$-pair,
because of the space-time overlap of the two hadronization processes, 
the two $W$'s cannot be treated as completely independent.
Consequently, one may see experimentally some effects
due to a ``cross-talk'' between the decays of the two $W$'s.
One of the possible effects (in addition to colour reconnection) 
could be a deformation of the hadron distributions
due to the ``coherence/interference'' effects in the hadronization process
called the Bose-Einstein effect.
This effect is of interest in itself, on the other hand it also can 
obscure the measurement
of the $W$ mass in the decay of $W$-pair using four-jet final states.
There are a variety of models, see \cite{Wmass-YR:1996}, describing 
the BE effect. In particular, the {\tt JETSET} package implements one of these 
models with the help of routine {\tt LuBoEi}.
The {\tt LuBoEi} routine introduces additional smearing of the momenta 
of the hadrons, 
after standard hadronization, so that the Bose-Einstein effect is reproduced.
In KoralW we include an example of the alternative model, 
see Ref.~\cite{JadachZalewski:1997},
where this additional smearing is done not by means of direct manipulation
of the hadron momenta, but with the help of an additional special weight.
In principle, this method is safer because it does not introduce spurious
long-range correlations of hadrons; it is therefore more relevant
to $W$-mass measurement from jet-jet effective masses than 
the {\tt LuBoEi} procedure.
The BE weight itself is programmed in the C++ class {\tt BEwtMaker}
in subdirectory {\tt B.E./src}.
(It uses the additional C++ functions hidden in the file {\tt partit1.C}.)
The code is rather compact, the bulk of the code is 
the construction of ``clusters''
of pions of the same sign, which are in some sense close to one another
in the phase space (their relative hyper-velocity is below certain 
maximum value). The subdirectories {\tt B.E./src} and {\tt B.E./fig}
contain the complete tool-box for analysing the
BE weight and finding out how big the BE effect is in the fitted $W$ mass.
For more details on this analysis of the BE effect programmed 
in the subdirectory {\tt B.E.}, see Ref.~\cite{JadachZalewski:1997}.

\subsection{Energy Distributions}

In this section we describe part of the code responsible for
the generation of the $s'$ photonic variable and decay channel type. 
The important observation is that the ISR photons and final-state
fermions are coupled solely by the $s'$ variable. 
This variable, on the one hand, 
determines the amount of energy radiated by the photons and at
the same time provides the effective CMS energy of the final fermions. 
Its generation is therefore an important step in the algorithm. 

As usual in the Monte Carlo algorithms, one can start from 
anything like the $s'$ distribution, at the end of the MC generation procedure,
this dummy distribution will be transformed into what we want and the original
one will get eliminated (by reweighting and/or rejecting MC events). 
The price for starting with a bad distribution is,
as usual, in the effectiveness of the program.
Since the cross-section varies by orders of magnitude with the 
energy and between various final states, the overhead in efficiency can be
really big.
For that reason we decided to pretabulate $s'$  distributions.
It is done for each final state separately and for the ``standard'' set of
cuts, as given in Sect. \ref{CUTS}. 

We see the potential disadvantage of this organization as well. 
In the case of cuts different from the ``standard'' ones, 
pretabulated spectra would gradually become inefficient. 
In our opinion, the gains due to the precise tune-up outweigh, however,  
the loses. 
In fact, one can always generate another set of files with new spectra,
if one is interested in very different cuts. 
We have created semi-automated tools for this task (not distributed with
KoralW~1.42) 
and encourage the interested user to contact us directly.

There is yet another trick that we used here. 
It seems natural that the pretabulated distributions are Born cross sections
(as a function of CMS energy). 
For the $s'$ generation, they will be
convoluted with the structure-function-type photonic density, with proper
soft singularity, and they are used directly for flavour generation.
However, the complicated four-fermion phase space leads, in some places, 
to events that are over-weighted  with respect to the global rejection weight. 
A global increase of the rejection weight would cause an unnecessary loss of 
effectiveness in the case of a run with constant weight. 
The pretabulation procedure distributions provide, as a by-product,
an efficient tool for compensating the above effect: 
with their help, we may provide the system of additional compensating weights,
depending on the flavours and the CMS energy.
In order to do it in a transparent way, we therefore
pretabulated the maximal weights, in addition to the total cross sections.
And in fact it is these maximal weights that the program uses for 
the generation of $s'$ at the crude level in the CC/NC-all mode. 
In the CC03 mode, the total cross section is used and the weights are 
well behaved.

We provide also an option for generating the distributions from the
user-supplied function. The function is called {\tt phot\_spec\_crud} and the
appropriate dip-switch {\tt i\_file=0} 
is located in {\tt karlud} routine. 
It is capable of reproducing the distributions used 
in the previous versions of KoralW.

At the technical level the organization is the following.
The pretabulated files, located in the directory {\tt data\_files}, are:
{\tt data\_wtmax.fit.smp1, data\_wtmax.fit.smp2} and 
{\tt data\_xsect.fit}.
They contain distributions of the maximal weight for two presamplers 
and the total cross section, respectively. 
They are tabulated as a function of the total energy for each
final state separately.
The tabulation is done in the energy range from 50 GeV to 250 GeV; beyond
these limits, a flat-distribution-matching value at 250 GeV is used.
The {\tt give\_phot\_spec\_crud} routine reads these files and stores
the data in some local internal variables. KoralW uses another set
of matrices with different  binning (currently the finer one): 
{\tt prob\_*\_*} (here {\tt *} denotes a wildcard for a part of 
a variable name), and a different energy range, 
from some {\tt emin} set to 1 GeV in the
{\tt korww} routine, to the {\tt emax=}$\sqrt{s}$. 
The arrays {\tt prob\_*\_*} contain channel-dependent distributions 
for the channel
choice as well as an inclusive distribution for the $s'$ generation.

\subsection{Semi-analytical Distributions}

The package for semi-analytical calculations distributed with KoralW
integrates the CC03 matrix element, based on the formulas of 
Ref.~\cite{muta:1986}. The calculation is done in massless approximation. 
Additional corrections due to NQCD, the non-diagonal CKM matrix and the 
Coulomb effect are included in the formulas. It is done exactly
as in the Monte Carlo generator, so the two nicely cross-check each
other. The Initial State Radiation
is also included. It is done in the LL approximation, by means of
the Structure Function formalism. The SFs are implemented up to the
third order with the YFS-type exponentiation and the 
non-leading YFS form factor \cite{jadach:1991, skrzypek:1992}. 
For the actual
calculations we provide a number of different approximations of these
SFs, as listed in Table \ref{tab:korwan-input}. For more details on the
integration method see Ref.~\cite{koralw:1995b}.

In the current version of KoralW
the semi-analytical part of the program
is enlarged with two functions {\tt s1wan(s1)} and
 {\tt s1s2wan(s1,s2)} for the one- and two-dimensional distributions of the
single and double $W$ invariant masses. These functions require
standard initialization of the {\tt korwan} routine.
Optionally, if the KorWan input parameter {\tt KeyMod} is increased by
10000 the calculations in KorWan are not executed 
and only the initialization is performed.

In addition we provide two simple applications of KorWan:
the average mass and average mass-loss calculations.
The average mass is defined as 
$(1/\sigma)\int dvds_1ds_2 (\sqrt{s_1}+\sqrt{s_2}-2M_W) d\sigma/(dvds_1ds_2)$, 
with $v$ being the photonic variable
$v=1-s'/s$ and $s_1, s_2$ being the invariant masses of the ``$W$-states''.
This calculation can be performed by invoking KorWan with
a negative value of the $s$-variable input parameter.
Alternatively, one can simply use the dedicated routine {\tt mavrg} with
arguments as given in Table \ref{tab:mavrg-input}.
The average mass loss is defined as 
$(\sqrt{s}/2)(1/\sigma)\int dv\; v\; d\sigma/dv$ with  $v=1-s'/s$. 
This calculation can be performed by invoking KorWan with the negative
{\tt KeyPho} input parameter.
Alternatively, one can simply use the dedicated routine {\tt mloss} with
arguments as given in Table \ref{tab:mloss-input}.

\section{Practical Use of the Program}

In this section we will familiarize the reader with the 
input and output parameters, and the usage of the present version of the
KoralW package.  
We will also present two simple demonstration main programs using 
the KoralW package.
Their double role is to serve as a useful template for the user 
to create his/her own main program and to help the user to check quickly that
the KoralW generators runs correctly.
We shall describe in detail all input parameters of KoralW.
We will also give a similar information on the semi-analytical
routine {\tt korwan}.

\subsection{Principal entries of KoralW}
The principal entries of the KoralW package, which the user has to call in
his/her application in order to generate series of the MC events, were
already listed and described briefly in Subsection~\ref{korww}.
Here we shall add more information on their functionality.
The calling sequence constituting a typical Monte Carlo run
will look as follows:
{\small
\begin{alltt}
 CALL KW\_ReaDataX('./data\_DEFAULTS',1,10000,xpar)  ! reading general defaults
 CALL KW\_ReaDataX('./user.input'    ,0,10000,xpar) ! reading user input
 CALL KW\_Initialize(xpar)                          ! initialize generator 
 DO loop=1,10000                                   ! loop over MC events
   CALL KW\_Make                                    ! generate single MC event
 ENDDO    
 CALL KW\_Finalize                                  ! final book-keeping, print
 CALL KW\_GetXSecMC(XSecMC,XErrMC)                  ! get total cross section
\end{alltt} 
}
\noindent
In the first call of {\tt KW\_ReaDataX}, default data is read into the array 
{\tt REAL*8 xpar(10000)}.
The KoralW has almost no data hidden in the source code.
(This is not true for TAUOLA and JETSET).
The file {\tt data\_DEFAULTS} is read first into array {\tt xpar}.
This file we provide in the distribution subdirectory {\tt data\_files}.
The user should {\em never modify it}. It can be copied to a local directory
or, better, a symbolic link should be created to the original 
{\tt data\_files/data\_DEFAULTS}.
The {\tt data\_DEFAULTS} is rather big and the user is usually interested
only in changing some subset of these data.
In the second call on  {\tt KW\_ReaDataX} the user can overwrite the default
data with his/her own smaller set on input data which are placed in the
{\tt user.input} file.
For example the simplest input data, which defines only CMS energy looks
as follows:
\begin{alltt}
BeginX
*<ia><----data-----><-------------------comments------------->
    1          190d0 CmsEne  =CMS total energy [GeV]
EndX
\end{alltt}
As we see, data cards start with the keyword {\tt BeginX} and end with  
the keyword {\tt EndX}.
The comment lines are allowed -- they start with {\tt *} in the first column.
The data themselves are in a fixed format, with the address $i$ in 
{\tt xpar(i)} followed by the data value and trailing comment.
The four examples of input data sets for the two demonstration programs 
{\tt KWdemo.f} and {\tt KWdemo2.f}
in the subdirectory {\tt demo.14x/190gev} provide useful templates 
for the typical user data.
The complete set of all user data in {\tt data\_DEFAULTS} is described in
very detail in Tables~\ref{tab:KoralW-input1}--\ref{tab:KoralW-input6}.
Obviously, the user is interested in manipulating only some of them
and will stick to default values in most of the cases.
The {\tt KW\_Initialize} is invoked to initialize the generator.
It reads input data from array {\tt xpar}, prints them and sends down
to various modules and auxiliary libraries.
The programs have to be called in strictly the same order
as in the above example.
At this point we are ready to generate series of the MC events.
The generation of a single event is done with the help of {\tt KW\_Make}.
After the generation loop is completed, we may invoke {\tt KW\_Finalize},
which does final book-keeping, prints various pieces of information 
on the MC run, and calculates the total MC integrated cross section 
in picobarns.
In order to obtain this cross section the user may call the routine
{\tt KW\_GetXSecMC(XSecMC,XErrMC)}.

\subsection{Input/Output Parameters}
As we have already explained in the previous section, the input
parameters enter through the {\tt xpar} array, being a parameter of 
{\tt CALL KW\_Initialize}.
The meaning of all of them is given in Tables 
\ref{tab:KoralW-input1}--\ref{tab:KoralW-input6}.

The principal output of KoralW is the Monte Carlo {\em event},
which is just a list of final-state four-momenta in [GeV] units and flavours,
encoded in the standard {\tt /hepevt/} common block.
At the present version we still provide REAL*4 version of 
the {\tt /hepevt/} common block.
If the user is interested in the parton momenta
before hadronization, then,
in addition to {\tt /hepevt/} common block,
they are also available through
{\small
\begin{alltt}
 CALL KW\_GetMomDec(p1,p2,p3,p4)     ! get momenta of four final fermions
 CALL KW\_GetBeams(q1,q2)            ! get beam momenta
 CALL KW\_GetPhotAll(NphAll,PhoAll)  ! get photon multiplicity and momenta
\end{alltt} 
}
\noindent
Alternatively, all the four-momenta are available via
the internal commons {\tt /momset/} and {\tt /momdec/}, see Tables 
\ref{tab:KoralW-momset} and \ref{tab:KoralW-MOMDEC} for details. 

For some special purposes, also the four-momenta in the effective CMS frame 
(used by default by the code for matrix elements calculation) 
are provided in the common block {\tt /cms\_eff\_momdec/}, as given in 
Table~\ref{tab:KoralW-CMS-eff-momdec}.
They may be useful for example to impose some additional cuts.
However, care must be taken, since such cuts might be {\it unphysical}.

In the case of a MC run with variable-weight events, the user is provided with
a main weight through
{\small
\begin{alltt}
 CALL KW\_GetWtMain(WtMain)              ! get main Monte Carlo weight
\end{alltt} 
}
\noindent
Of course, for constant-weight runs {\tt WtMain = 1}.
For special purposes the user may be also interested in auxiliary weights,
which are provided with the help of
{\small
\begin{alltt}
 CALL KW\_GetWtAll(WtMain,WtCrud,WtSet)  ! get all Monte Carlo weights
\end{alltt} 
}
\noindent
where {\tt REAL*8 WtSet(100)} is an array of weights described
in Table \ref{tab:KoralW-weights}.
Alternatively these weights are available from
the common block {\tt /wtset/}, see Table \ref{tab:KoralW-weights}.
The auxiliary weight should be defined as {\tt WtCrud*WtSet(i)},
and the corresponding integrated  cross section simply obtained by
multiplying its average by
the crude {\em normalization} cross section, which is provided through
{\small
\begin{alltt}
 CALL KW\_GetXSecNR(XSecNR,XErrNR)       ! get normalization x-section
\end{alltt} 
}
\noindent
The complete description of post-generation
output parameters from {\tt KW\_Finalize} is collected in 
Table~\ref{tab:KoralW-output1}. 

Finally, the description of input-output
parameters of the semi-analytical routine {\tt korwan} is given in 
Table~\ref{tab:korwan-input}. It should be recalled here, that the
semi-analytical functions {\tt s1wan} and {\tt s1s2wan} 
are initialized {\it by} the {\tt korwan}
routine. To this end, for pure initialization, 
a special setting of its {\tt KeyMod} argument, {\tt
KeyMod > 10000}, is provided. Tables~\ref{tab:mavrg-input} and 
\ref{tab:mloss-input} describe the input/output of the 
{\tt mavrg} and {\tt mloss} routines.

\subsection{Printouts of the Program}
\label{PRINTOUTS}

In this section we describe a printout of the demonstration program
{\tt KWdemo} in the all-four-fermion mode, shown in Appendix C.
The printout starts with the detailed specification of the actually used input
parameters. Also, logos of all the activated libraries (TAUOLA, PHOTOS,
JETSET) are printed here.
Next, the printout of one full event (in the standard PDG convention)
is shown. 
The final reports of KoralW are collected in four windows: $A$, $B$, $C$ 
and $D$. 

Window $A$ provides a technical internal report of the {\tt karlud} routine,
i.e.~information on the crude distribution and the Born matrix element.
The line $a0$ gives the total number of generated {\it weighted}
events. The line $a1$ shows the number of events with the negative weight
{\tt wtcrud}. This entry should be equal to zero. The line $a2$ provides the
value of the master crude distribution.  The line $a3$ shows the
average of the weight {\tt wtcrud}, and the line $a4$ -- the corresponding
``cross section'' 
(no matrix element is included here -- only the crude distribution formula).
The lines $a5$-$a8$ repeat the information of
the lines $a0$-$a1$ and $a3$-$a4$,
but for the weight {\it with} the Born matrix element, i.e.\
{\tt wtcrud*wtset(1)}. The lines $a9$-$a11$ provide information on how many of
the events have the weight {\tt wtcrud*wtset(1)} over the {\tt wtmax},
the maximal weight for rejection. Their contribution to the cross
section in absolute units (picobarns) and relative to the Born cross section
is listed in $a10$ and $a11$, respectively.
Note that there are no $\bbeta$-functions included in any of
the printouts in Window $A$.

Window $B$ is devoted to the technical information on the QED radiative
corrections in different orders in $\alpha$ within the YFS framework. 
This information is quite important, 
because it shows how big the contributions 
of subsequent orders of the perturbative series are, and thus allows us to
estimate the missing higher-order effects. The entries $b3$-$b6$ provide
the values of the total cross section (in picobarns) in the orders
${\cal O}(\alpha^0)_{exp}$-${\cal O}(\alpha^3)_{exp}$, 
while the entries  $b17$-$b19$
give the differences of these values in subsequent orders.
The entries $b7$-$b16$ contain the information on the contributions
from the individual YFS residuals $\bbeta_i$, also in different
orders in $\alpha$. The differences between various residuals and
various orders for a given residual are provided in the entries $b20$-$b25$. 

Window $C$ is the most important from the user's point of view.
It provides the total number of generated events ($c1$),
the number of accepted events ($c2$), 
the best-order total cross section (in picobarns) 
with the absolute error ($c3$),
as well as the relative error ($c4$), for the generated statistics sample. 
Then the information on the negative-weight events is given as a number
of such events ($c5$) and their relative contribution to the cross section 
($c6$), followed by similar information on the overweighted events 
($c7$ and $c8$, respectively). All these four numbers should be as small
as possible. 

Window $D$ contains some supplementary information and is printed out only
if the 4-fermion matrix element is included in the calculation.
Part~I provides the values of the average weight ($d1$) and
the total cross section ($d2$) for the $WW$ process (if such exist
for a given final state). Part~II contains some additional
information on the 4-fermion process, as:
the average Born level weight ($d3$), the average
total weight ($d4$), the total cross section ($d5$) --
this should be the same as in Window~$C$ (entry $c3$), 
the relative contribution to the cross section of the non-$WW$
diagrams, i.e. the size of the background to the $W$-pair production
($d6$, calculated from entries $d2$ and $d5$, and $d7$, calculated
from a dedicated weight, monitored during the event generation).

Finally, the report on different channels is printed out. It includes
the codes of the decay $W$ and/or $Z$ channels (first column),
the flavours of the final-state fermions in the ``human-readable'' format
(second column),  the values of the total cross section and its
absolute error (in pb) for individuals channels (third column),
the ratio of the maximum weight over the average weight (fourth column) 
and over the average non-zero weight (fifth column), the fraction
of events generated for a given channel with respect to the total
number of events (sixth column) and a similar fraction for non-zero-weight 
events (last column). The last line of this report provides
the value of the total cross section and its error (in pb) summed over
all open channels.    
  
This completes the description of the output of KoralW. The
remaining entries shown in the demo output are produced by the demo
main program.

\subsection{Random Number Generators}

The KoralW code uses exclusively one of three random number generators:
{\tt RANMAR}, {\tt ECURAN} or {\tt CARRAN}. These single precision
generators are called by one double precision interface routine {\tt
VARRAN}. The choice between generators is done with the help of the key
{\tt KeyRnd} ($1=${\tt RANMAR}, $2=${\tt ECURAN}, $3=${\tt CARRAN}). 
In order to avoid possible interference with libraries, {\tt RANMAR} 
of KoralW is renamed to {\tt MARRAN}. 
JETSET, PHOTOS and TAUOLA have their own independent 
random number generators.

\section*{Acknowledgements}
We thank the CERN TH and EP Divisions and all four LEP Collaborations 
for their support. Two of us (W.~P. and M.~S.) also thank 
Prof.~Lee L.~Riedinger for the support and kind hospitality 
of the Department of Physics and Astronomy
of the University of Tennessee, Knoxville, TN, 
where part of this work was done. 
Two of us (M.~S. and Z.~W.) would like to thank Prof.~W.~Hollik 
for the hospitality of ITP, Karlsruhe Universit\"at. 
We would like to express our gratitude to S.~Jezequel, M.~Gr\"unewald and 
M.~Witek for their help in testing the program and valuable comments.

\newpage
\section*{Appendix A: \\ Program Parameters and Their Settings}

\begin{table}[h]
\centering
\begin{small}
\begin{tabular}{|l|p{13.0cm}|}
\hline
Parameter & Position and meaning  \\ 
\hline\hline
{\tt cmsene}        & {\tt xpar(1) (=180) }: $\sqrt{s}$, centre-of-mass (CMS) energy  [GeV]\\
{\tt gmu}           & {\tt xpar(2) (=1.16639d-5) }: $G_F$, Fermi constant in [GeV]\\
{\tt alfwin}        & {\tt xpar(3) (=128.07d0)}:
                           $1/\alpha_W$  inverse QED coupling constant at $M_W$ scale\\
{\tt amaz}          & {\tt xpar(4) (=91.1888)}: $M_Z$, mass  of $Z$ boson [GeV]\\
{\tt gammz}         & {\tt xpar(5) (=2.4974)}: $\Gamma_Z$, width of $Z$ boson in [GeV]\\
{\tt amaw}          & {\tt xpar(6) (=80.230)}: $M_W$, mass  of $W$ boson [GeV]\\
{\tt gammw}         & {\tt xpar(7) (=-2.03)}:  $\Gamma_W$, width of $W$ boson [GeV],
                           for {(\tt gammw < 0)} $\Gamma_W$ is recalculated from $G_\mu$, $M_W$ and $\alpha_S$\\
{\tt vvmin}         & {\tt xpar(8) (=1d-6}): Minimum $v$-variable (dimensionless), 
                                             infra-red cut-off\\
{\tt vvmax}         & {\tt xpar(9) (=0.99)}: Maximum value of $v$-variable\\
{\tt wtmax}         & {\tt xpar(10) (=-1)}: Maximum weight for rejection,  
                           for {\tt wtmax < 0} redefined inside program\\
{\tt amh}           & {\tt xpar(11) (=1000)}: Higgs mass [GeV]\\
{\tt agh}           & {\tt xpar(12) (=1)}:    Higgs width [GeV]\\
{\tt alpha\_s}      & {\tt xpar(13) (=0.12)}: QCD coupling constant\\
{\tt arbitr}        & {\tt xpar(14) (=600)}:  minimum visible $p_T$ [GeV$^2$]\\
{\tt arbitr1}       & {\tt xpar(15) (=8)}: invariant-mass cut for $e^+e^-f\bar{f}$ [GeV$^2$]\\
{\tt themin}        & {\tt xpar(16) (=1d-6)}: minimum $\theta$ [rad] relative to beam (0=no cut)\\
{\tt arbitr2}       & {\tt xpar(17) (=300)}: maximum $p_T$ of photons 
                      in $e^+e^-f\bar{f}$ channels [GeV$^2$] 
                      (if {\tt arbitr2} $\leq 0$ then no cut)\\
{\tt wtmax\_cc03}   & {\tt xpar(18) (=-1)}: maximum CC03 weight for rejection in {\tt KeySmp=0},
                           for {\tt wtmax < 0} redefined inside program\\
{\tt PReco}         & {\tt xpar(19) (=0)}: Colour Reconnection Probability\\
\hline
\end{tabular}
\end{small}
\caption{\it List of input parameters of the {\tt KoralW} generator
             in {\tt xpar} vector. Default values in brackets.}
\label{tab:KoralW-input1}
\end{table}

\begin{table}
\centering
\begin{small}
\begin{tabular}{|l|p{13.0cm}|}
\hline
Parameter & Position and meaning  \\ 
\hline\hline
                   & {\tt xpar(21-57)}: Values of TGCs: set 1, most general set -- 
                     complex numbers, default values are wild random, not shown\\
{\tt g1(1)   }     & {\tt =DCMPLX(xpar(21),xpar(31))} $=g_1^z$,     for $WWZ$ vertex \\
{\tt kap(1)  }     & {\tt =DCMPLX(xpar(22),xpar(32))} $=\kappa_z$,  for $WWZ$ vertex \\
{\tt lam(1)  }     & {\tt =DCMPLX(xpar(23),xpar(33))} $=\lambda_z$, for $WWZ$ vertex \\
{\tt g4(1)   }     & {\tt =DCMPLX(xpar(24),xpar(34))} $=g_4^z$,     for $WWZ$ vertex \\
{\tt g5(1)   }     & {\tt =DCMPLX(xpar(25),xpar(35))} $=g_5^z$,     for $WWZ$ vertex \\
{\tt kapt(1) }     & {\tt =DCMPLX(xpar(26),xpar(36))} $=\tilde\kappa_z$,  for $WWZ$ vertex \\
{\tt lamt(1) }     & {\tt =DCMPLX(xpar(27),xpar(37))} $=\tilde\lambda_z$, for $WWZ$ vertex\\
{\tt g1(2)   }     & {\tt =DCMPLX(xpar(41),xpar(51))} $=g_1^g$,     for $WW\gamma$ vertex\\
{\tt kap(2)  }     & {\tt =DCMPLX(xpar(42),xpar(52))} $=\kappa_g$,  for $WW\gamma$ vertex\\
{\tt lam(2)  }     & {\tt =DCMPLX(xpar(43),xpar(53))} $=\lambda_g$, for $WW\gamma$ vertex\\
{\tt g4(2)   }     & {\tt =DCMPLX(xpar(44),xpar(54))} $=g_4^g$,     for $WW\gamma$ vertex\\
{\tt g5(2)   }     & {\tt =DCMPLX(xpar(45),xpar(55))} $=g_5^g$,     for $WW\gamma$ vertex\\
{\tt kapt(2) }     & {\tt =DCMPLX(xpar(46),xpar(56))} $=\tilde\kappa_g$,   for $WW\gamma$ vertex\\
{\tt lamt(2) }     & {\tt =DCMPLX(xpar(47),xpar(57))} $=\tilde\lambda_g$,  for $WW\gamma$ vertex\\
                   & {\tt xpar(61-65)}:  Values of TGCs: set 2,
                                     see CERN 96-01, Vol. 1, p. 525 \\
{\tt  delta\_Z }   & {\tt =xpar(61)} $=\delta_Z$ \\ 
{\tt  x\_gamma }   & {\tt =xpar(62)} $=x_{\gamma}$ \\
{\tt  x\_Z     }   & {\tt =xpar(63)} $=x_Z$ \\
{\tt  y\_gamma }   & {\tt =xpar(64)} $=y_{\gamma}$ \\
{\tt  y\_Z     }   & {\tt =xpar(65)} $=y_Z$\\
                   & {\tt xpar(71-73)}: Values of TGCs: set 3,
                                     see CERN 96-01, Vol. 1, p. 525\\
{\tt alpha\_Wphi}  & {\tt =xpar(71)} $=\alpha_{W\phi}$\\
{\tt alpha\_Bphi}  & {\tt =xpar(72)} $=\alpha_{B\phi}$\\
{\tt alpha\_W   }  & {\tt =xpar(73)} $=\alpha_W$\\
\hline
\end{tabular}
\end{small}
\caption{\it List of input parameters of the {\tt KoralW} generator
             in {\tt xpar} vector (cont.). Default values in brackets.}
\label{tab:KoralW-input2}
\end{table}

\begin{table}
\centering
\begin{small}
\begin{tabular}{|l|p{13.0cm}|}
\hline
Variable & Position and meaning  \\ 
\hline\hline
{\tt KeyISR}  &  {\tt xpar(1011) (=1) }\\
              &  {\tt =0}   Initial State Radiation is OFF\\
              &  {\tt =1}   Initial State Radiation is ON\\
{\tt KeyFSR}  &  {\tt xpar(1012) (=0)}  Final State Radiation switch, INACTIVE\\
{\tt KeyNLL}  &  {\tt xpar(1013) (=1)}\\
              &  {\tt =0} sets to zero the Next-to Leading $\alpha/\pi$ terms in
                 the YFS form factor, useful for comparisons\\
              &  {\tt =1} the $\alpha/\pi$ terms are kept in the YFS form factor\\
{\tt KeyCul}  &  {\tt xpar(1014) (=1)}\\
              &  {\tt =0} Coulomb correction is OFF\\
              &  {\tt =1} Coulomb correction is ON\\
{\tt KeyBra}  &  {\tt xpar(1021) (=1) }\\
              &  sets $W$ branching ratios, used for normalization of CC03 matrix element only\\
              &  {\tt =0} Born values (no mixing) with Naive QCD
                 (if $\alpha_S=0$: $ud, cs = 1/3$,\; $e\nu,\mu\nu,\tau\nu=1/9$,\; others $=0$)\\
              &  {\tt =1} with CKM mixing and Naive QCD, defined in {\tt KW\_Initialize}\\
              &  {\tt =2} with CKM mixing and Naive QCD,
                          calculated in IBA from the CKM matrix (PDG '98)\\
{\tt KeyMas}  &  {\tt xpar(1022) (=1) }\\
              &  {\tt =0} Massless kinematics; $\tau$ decay, radiative corrections in decay
                          and hadronization must be switched off\\
              &  {\tt =1} Massive kinematics, masses of fermions are read from {\tt xpar}\\
{\tt KeyZet}  &  {\tt xpar(1023) (=0) }\\
              &  {\tt =0} $Z$ width in $Z$ propagator: $(s/M_Z) \Gamma_Z$\\
              &  {\tt =1} $Z$ width in $Z$ propagator:   $M_Z \Gamma_Z$\\
              &  {\tt =2} no $Z$ width in $Z$ propagator\\
{\tt KeySpn}  &  {\tt xpar(1024) (=1) }\\
              &  {\tt =0} spin effects are switched OFF in $W$ decays, for tests only\\
              &  {\tt =1} spin effects are switched ON in $W$ decays\\
{\tt KeyRed}  &  {\tt xpar(1025) (=0) }\\
              &  {\tt =0} ``sophisticated'' reduction of massive  to massless four-vectors 
                          for CC03 internal Matrix Element\\
              &  {\tt =1} ``brute-force'' reduction of massive to massless four-vectors 
                          for CC03 internal Matrix Element, four-momentum NON-conserving\\
              &  {\tt =2} no reduction at all of massive to massless four-vectors 
                          for CC03 internal Matrix Element\\
{\tt KeyWu}   &  {\tt xpar(1026) (=0) }\\
              &  {\tt =0} $W$ width in $W$ propagator: $(s/M_W) \Gamma_W$\\
              &  {\tt =1} $W$ width in $W$ propagator: $M_W \Gamma_W$\\
              &  {\tt =2} no $W$ width in $W$ propagator\\
\hline
\end{tabular}
\end{small}
\caption{\it List of input parameters of the {\tt KoralW} generator 
             in {\tt xpar} vector (cont.). Default values in brackets.}
\label{tab:KoralW-input3}
\end{table}

\begin{table}
\centering
\begin{small}
\begin{tabular}{|l|p{13.0cm}|}
\hline
Parameter & Position and meaning  \\ 
\hline\hline
{\tt KeyWgt}  &  {\tt xpar(1031) (=0) }\\
              &  {\tt =0} constant weight {\tt wtmod}=1, for apparatus Monte Carlo\\
              &  {\tt =1} variable weight {\tt wtmod} events\\
              &  {\tt =2} for special purposes: {\tt wtmod}=1 for internal matrix element, 
                 AND varying weight for external matrix element\\
{\tt KeyRnd}  &  {\tt xpar(1032) (=1) }\\
              &  {\tt =1} RANMAR random number generator\\ 
              &  {\tt =2} ECURAN random number generator\\ 
              &  {\tt =3} CARRAN random number generator\\ 
{\tt KeySmp}  &  {\tt xpar(1033) (=2) }\\
              &  {\tt =0} presampler set as in KoralW v. 1.02-1.2x, i.e. CC03 oriented\\ 
              &  {\tt =1} first presampler for all 4-fermion final states\\ 
              &  {\tt =2} second presampler for all 4-fermion final states\\ 
              &  {\tt =3} 50/50 mixed (1+2) presampler for all 4-fermion final states\\
{\tt KeyMix}  &  {\tt xpar(1041) (=0) }\\
              &  {\tt =0} ``renormalization scheme'' of LEP2 Workshop (recommended)\\
              &  {\tt =1} ``renormalization scheme'' based on $G_\mu$\\
              &  {\tt =2} ``bare Born'' (for tests)\\
{\tt Key4f}   &  {\tt xpar(1042) (=1) }\\
              &  {\tt =0} External Matrix Element OFF\\
              &  {\tt =1} External Matrix Element ON\\
{\tt KeyAcc}  &  {\tt xpar(1043) (=0) }\\
              &  {\tt =0} anomalous $WWV$ couplings in internal CC03 matrix element OFF\\
              &  {\tt >0} anomalous $WWV$ couplings in internal CC03 matrix element ON\\
              &  {\tt =1} the most general (complex number) TGCs in the notation
                        of K.~Hagiwara et al., Nucl. Phys. {\bf B282} (1987) 253\\
              &  {\tt =2} parametrization of CERN 96-01,
                        Vol. 1, p. 525: $\delta_Z, x_\gamma, x_Z, y_\gamma, y_Z$\\
              &  {\tt =3} parametrization of CERN 96-01,
                        Vol. 1, p. 525: $\alpha_{W\phi}, \alpha_{B\phi}, \alpha_W$\\
{\tt KeyZon}  &  {\tt xpar(1044) (=1) }\\
              &  {\tt KeyZon=0} ZZ type final states OFF \\
              &  {\tt KeyZon=1} ZZ type final states ON \\
{\tt KeyWon}  &  {\tt xpar(1045) (=1) }\\
              &  {\tt KeyWon=0} WW type final states OFF \\
              &  {\tt KeyWon=1} WW type final states ON \\
\hline
\end{tabular}
\end{small}
\caption{\it List of input parameters of the {\tt KoralW} generator
             in {\tt xpar} vector (cont.). Default values in brackets.}
\label{tab:KoralW-input4}
\end{table}

\begin{table}
\centering
\begin{small}
\begin{tabular}{|l|p{13.0cm}|}
\hline
Parameter & Position and meaning  \\ 
\hline\hline
{\tt KeyDwm}    &  {\tt xpar(1055) (=0) }\\
                &  Sets decay channel of $W^-$ or ``$Z_1$'' resonance,
                   depending on {\tt KeyWon/KeyZon}\\
                &  {\tt =0} inclusive, otherwise exclusive modes for $W$ are\\
                &  $\begin{bmatrix}  =1  &  2 &  3 &  4 &  5 &  6 &    7 &      8 &       9\\ 
                                      ud & cd & us & cs & ub & cb & e\nu & \mu\nu & \tau\nu 
                    \end{bmatrix}$ \\
                &  and for $Z$ are \\
                &  $\begin{bmatrix}  = 1&  2&  3&  4&  5&  6&      7&        8&   \\ 
                                      dd& uu& ss& cc& bb& ee& \mu\mu& \tau\tau& 
                     \end{bmatrix}$
                     $\begin{bmatrix}        =9&             10&              11\\
                                  \nu_e\nu_e& \nu_\mu\nu_\mu& \nu_\tau\nu_\tau 
                     \end{bmatrix}$ \\
{\tt KeyDwp}    &  {\tt xpar(1056) (=0) }\\
                &  Sets decay channel of $W^+$ or ``$Z_2$'' resonance, depending 
                   on {\tt KeyWon/KeyZon}, similar assignments as for $W^-$ or ``$Z_1$''\\
                &  Note: In the inclusive assignment {\tt KeyDwm=0, KeyDwp=0} 
                   the auxiliary user mask {\tt Umask } is used by the program.
                   It allows the user to configure arbitrary menu of {\tt KeyDwm, KeyDwp}.
                   The {\tt Umask} is passed through {\tt xpar(1101)-xpar(1302)} parameters.
                   Final states that may come from either $WW$ or $ZZ$ 
                   are classified as $WW$-type except for
                   $ussu, ubbu, cddc$ and $cbbc$, which are classified as
                   $ZZ$-type. See Appendix B for more details.\\
{\tt Nout}      &  {\tt xpar(1057) (=-1) }\\
                &   Output unit number for the generator (if $<0$ then {\tt Nout=16})\\
{\tt Jak1}      &  {\tt xpar(1071) (=0) }\\
                &  Input for TAUOLA, defines decay mode of $\tau^+$ in $W^+$ decay\\
{\tt Jak2}      &  {\tt xpar(1072) (=0) }\\
                &  Input for TAUOLA, defines decay mode of $\tau^-$ in $W^-$ decay\\
                &  {\tt Jak1,Jak2 = -1} TAUOLA is switched OFF\\
                &  {\tt Jak1,Jak2 = 0} requests all $\tau^\pm$ decay channels to be simulated\\ 
                &  {\tt Jak1,Jak2 > 0} single specific $\tau^\pm$ decay channel, see TAUOLA manual\\
{\tt itdkrc}    &  {\tt xpar(1073) (=1) }\\
                &  Input  for TAUOLA, radiative corrections in leptonic $\tau$ decays switch\\
                &  {\tt itdkrc=1} corrections are  ON\\
                &  {\tt itdkrc=0} corrections are  OFF\\
{\tt ifphot}    &  {\tt xpar(1074) (=1) }: PHOTOS, activation switch\\
                &   {\tt =1} PHOTOS is ON\\
                &   {\tt =0} PHOTOS is OFF\\
{\tt ifhadM}    &  {\tt xpar(1075) (=1) }: $W^-$ hadronization activation switch (JETSET)\\
{\tt ifhadP}    &  {\tt xpar(1075) (=1) }: $W^+$ hadronization activation switch (JETSET)\\
                &  {\tt ifhadm,ifhadp=1} hadronization is ON\\
                &  {\tt ifhadm,ifhadp=0} hadronization is OFF\\
                &  In the present version {\tt ifhadm} and {\tt ifhadp} have to be equal!\\
{\tt Umask}     &  {\tt xpar(1101-1302) }\\
                &   user supplied {\tt Umask} to activate a specific menu of the final states.
\\
\hline
\end{tabular}
\end{small}
\caption{\it List of input parameters of the {\tt KoralW} generator
             in {\tt xpar} vector (cont.). Default values in brackets.}
\label{tab:KoralW-input5}
\end{table}

\begin{table}
\centering
\begin{small}
\begin{tabular}{|l|p{13.0cm}|}
\hline
Parameter & Meaning  \\ 
\hline\hline
{\tt amel}          &  {\tt xpar(100) (=0.51099906d-3) }  
                       Electron mass used by the bremsstrahlung generator, reset to {\tt amafin(11)}\\
{\tt alfinv}        &  {\tt xpar(101) (=137.0359895) } 
                      $1/\alpha(0)$ Inverse QED coupling constant at Thomson scale $Q^2=$0\\
{\tt gpicob}        &  {\tt xpar(102) (=389.37966d6) }  
                       Conversion factor from GeV$^{-2}$ to pb\\
{\tt br(1:20)}      &  {\tt xpar(131-139)}
                       $W$ branching ratios; numbering of entries is:
                       $1=ud, 2=cd, 3=us, 4=cs, 5=ub, 6=cb, 7=e, 8=\mu, 9=\tau$\\
{\tt amafin(20)}    &  {\tt xpar(500 +10*KF +6)}
                       Masses of the $W$ decay products; used entries KF (in PDG notation) are:
                       $ 1=d,  2=u,  3=s,  4=c,  5=b, 6=t$,
                       $11=e, 12=\nu_e, 13=\mu, 14=\nu_\mu, 15=\tau, 16=\nu_\tau$;
                       masses of $\tau$ and $\nu_\tau$ have to be independently set to the same
                       numerical values in intialization of TAUOLA\\
{\tt vckm(1:3,1:3)} &  {\tt xpar(111-119)}
                       CKM matrix elements\\
\hline
\end{tabular}
\end{small}
\caption{\it List of input parameters of the {\tt KoralW} generator
             in {\tt xpar} vector (cont.). Default values in brackets.}
\label{tab:KoralW-input6}
\end{table}

\begin{table}
\centering
\begin{small}
\begin{tabular}{|l|p{13.0cm}|}
\hline
Parameter & Meaning  \\ \hline\hline
{\tt qeff1(4)}     & Effective parameter for matrix element ($e^-$ effective beam, CMS)\\
{\tt qeff2(4)}     & Effective parameter for matrix element ($e^+$ effective beam, CMS)\\
{\tt sphum(4)}     & Sum of four-momenta of ISR photons, CMS\\
{\tt sphot(100,4)} & Four-momenta of ISR photons, CMS\\
{\tt nphot}        & Multiplicity of ISR photons\\
\hline
\end{tabular}
\end{small}
\caption{\it List of four-momenta in the internal common block {\tt /momset/}
             of the {\tt KoralW} generator.}
\label{tab:KoralW-momset}
\end{table}

\begin{table}
\centering
\begin{small}
\begin{tabular}{|l|p{13.0cm}|}
\hline
Parameter & Meaning  \\
\hline\hline
{\tt q1(4)} & Four-momentum of the $W^-/Z_1$ resonance, $=p_1+p_2$, in CMS\\
{\tt q2(4)} & Four-momentum of the $W^+/Z_2$ resonance, $=p_3+p_4$, in CMS\\
{\tt p1(4)} & Four-momentum of the fermion from $W^-/Z_1$ decay, CMS\\
{\tt p2(4)} & Four-momentum of the antifermion from $W^-/Z_1$ decay, CMS\\
{\tt p3(4)} & Four-momentum of the fermion from $W^+/Z_2$ decay, CMS\\
{\tt p4(4)} & Four-momentum of the antifermion from $W^+/Z_2$ decay, CMS\\
\hline
\end{tabular}
\end{small}
\caption{\it List of four-momenta in the internal common block {\tt /momdec/}
             of the {\tt KoralW} generator.}
\label{tab:KoralW-MOMDEC}
\end{table}

\begin{table}
\centering
\begin{small}
\begin{tabular}{|l|p{13.0cm}|}
\hline
Parameter & Meaning  \\ 
\hline\hline
{\tt effbeam1(4)} & Four-momentum of the $e^-$ beam, along $z$-axis, in effective CMS\\
{\tt effbeam2(4)} & Four-momentum of the $e^+$ beam, in effective CMS\\
{\tt effp1(4)}    & Four-momentum of the fermion from $W^-/Z_1$ decay, effective CMS\\
{\tt effp2(4)}    & Four-momentum of the antifermion from $W^-/Z_1$ decay, effective CMS\\
{\tt effp3(4)}    & Four-momentum of the fermion from $W^+/Z_2$ decay, effective CMS\\
{\tt effp4(4)}    & Four-momentum of the antifermion from $W^+/Z_2$ decay, effective CMS\\
\hline
\end{tabular}
\end{small}
\caption{\it List of four-momenta in the internal common block {\tt /cms\_eff\_momdec/} 
             of the {\tt KoralW} generator.}
\label{tab:KoralW-CMS-eff-momdec}
\end{table}

\begin{table}
\centering
\begin{small}
\begin{tabular}{|l|p{11.0cm}|}
\hline
Parameter & Meaning  \\ 
\hline\hline
{\tt wtcrud}       & Crude weight, necessary to build the total weight out of {\tt wtset}\\
{\tt wtmod}        & Best weight \\
                   & For {\tt KeyISR=0 wtmod=wtcrud*wtset(1)},\\
                   & For {\tt KeyISR=1 wtmod=wtcrud*wtset(4)}\\
{\tt wtset(1-100)} & Born matrix element with various $\bbeta$ contributions --
                      to get total weight must be multiplied by {\tt wtcrud}\\
{\tt wtset(1)}     & Zero-order complete ($\bbeta_0$)\\
{\tt wtset(2)}     & First-order complete ($\bbeta_0 +\bbeta_1$)\\
{\tt wtset(3)}     & Second-order complete ($\bbeta_0  +\bbeta_1 +\bbeta_2$)\\
{\tt wtset(4)}     & Third-order complete ($\bbeta_0  +\bbeta_1 +\bbeta_2 +\bbeta_3$)\\
{\tt wtset(10)}    & ${\cal O}(\alpha^0)$  contribution \\
{\tt wtset(11)}    & ${\cal O}(\alpha^1)$  0-real, 1-virtual photon contribution\\
{\tt wtset(12)}    & ${\cal O}(\alpha^1)$  1-real, 0-virtual photon contribution\\
{\tt wtset(13)}    & ${\cal O}(\alpha^2)$  0-real, 2-virtual photon contribution\\
{\tt wtset(14)}    & ${\cal O}(\alpha^2)$  1-real, 1-virtual photon contribution\\
{\tt wtset(15)}    & ${\cal O}(\alpha^2)$  2-real, 0-virtual photon contribution\\
{\tt wtset(16)}    & ${\cal O}(\alpha^3)$  0-real, 3-virtual photon contribution\\
{\tt wtset(17)}    & ${\cal O}(\alpha^3)$  1-real, 2-virtual photon contribution\\
{\tt wtset(18)}    & ${\cal O}(\alpha^3)$  2-real, 1-virtual photon contribution\\
{\tt wtset(19)}    & ${\cal O}(\alpha^3)$  3-real, 0-virtual photon contribution\\
{\tt wtset(40)}    & {\tt wtmod4f}, external matrix element \\
{\tt wtset(41-49)} & wt4f(1-9) additional weights from ext.\
                     matrix element, if provided\\
\hline
\end{tabular}
\end{small}
\caption{\it List of output weights in the common block {\tt /wgtall/}
             of the {\tt KoralW} generator.}
\label{tab:KoralW-weights}
\end{table}

\begin{table}
\centering
\begin{small}
\begin{tabular}{|l|p{10.0cm}|}
\hline
Parameter & Meaning  \\ 
\hline\hline
{\tt XSecMC}               & {\sl Principal} best total Monte Carlo cross section [pb]\\
                           & For {\tt KeyISR=0} Born cross section\\
                           & For {\tt KeyISR=1} third-order exponentiated cross section\\
{\tt XErrMC}               & Its absolute error [pb]\\
{\tt XSecNR}               & Normalization cross section: \\
                           & For {\tt KeyWgt=0} principal cross section {\tt XSecMC} [pb],\\
                           & For {\tt KeyWgt=1} crude cross section {\tt XCrude} [pb]\\
{\tt XErrNR}               & Its absolute error [pb]\\
{\tt NevMC}                & Total number of generated Monte Carlo events\\
\hline
\end{tabular}
\end{small}
\caption{\it List of output parameters of KoralW generator transferred through parameters
             of the getter type subprograms {\tt KW\_GetXSecMC(XSecMC,XErrMC)},  
             {\tt KW\_GetNevMC(NevMC)} and {\tt KW\_GetXSecNR(XSecNR,XErrNR)}.}
\label{tab:KoralW-output1}
\end{table}

\begin{table}
\centering
\begin{small}
\begin{tabular}{|l|p{11.0cm}|}
\hline
Parameter & Meaning  \\ \hline\hline
               &  INPUT \\
\hline
{\tt svar}     &  $s$, CMS energy squared [GeV$^2$]\\
{\tt keypho}   &  As {\tt KeyMod} of {\tt korwan}\\
{\tt kaccbre}  &  As {\tt KeyPre} of {\tt korwan}\\
\hline
               &  OUTPUT \\
\hline
{\tt dmavrg}   &  Mass average  in GeV\\
{\tt dmerr}    &  Absolute error in GeV\\
\hline
\end{tabular}
\end{small}
\caption{\it List of arguments of the {\tt mavrg} routine. }
\label{tab:mavrg-input}
\end{table}

\begin{table}
\centering
\begin{small}
\begin{tabular}{|l|p{11.0cm}|}
\hline
Parameter & Meaning  \\ \hline\hline
               &  INPUT\\
\hline
{\tt svar}     &  $s$, CMS energy squared [GeV$^2$]\\
{\tt keypho}   &  As {\tt KeyMod} of {\tt korwan}\\
{\tt eeps}& Absolute precision of integration\\
\hline
               &  OUTPUT\\
\hline
{\tt vvloss}   &  Average mass loss  in GeV\\
{\tt vverr}    &  Absolute error in GeV\\
\hline
\end{tabular}
\end{small}
\caption{\it List of arguments of the {\tt mloss} routine. }
\label{tab:mloss-input}
\end{table}

\begin{table}
\centering
\begin{small}
\begin{tabular}{|l|p{13.0cm}|}
\hline
Parameter & Meaning  \\ \hline\hline
&  INPUT
\\
\hline
{\tt svar}     &  $s$, CMS energy squared [GeV$^2$]\\
{\tt vvmin}    &  $v_{min}$, minimal $v$ variable, in most cases should be set to 0\\
{\tt vvmax}    &  $v_{max}$, maximal $v$ variable\\
{\tt keymod}   &  Defines type of structure functions used for ISR:\\
               &  {\tt =~~0} No ISR, Born,\\
               &  {\tt =300} Zero   Order, YFS style,\\
               &  {\tt =301} First  Order, YFS style,\\
               &  {\tt =302} Second Order, YFS style,\\
               &  {\tt =303} Third  Order, YFS style,\\
               &  {\tt =502} Second Order, Gribov-Kuraev-Fadin style,\\
               &  {\tt =310} First-Order YFS Beta0 only,\\
               &  {\tt =311} First-Order YFS Beta1 only,\\
               &  {\tt =320} Second-Order YFS Beta0 only,\\
               &  {\tt =321} Second-Order YFS Beta1 only,\\
               &  {\tt =322} Second-Order YFS Beta2 only,\\
               &  {\tt <0} as {\tt (-keymod)} but multiplied by $v$ 
                  differential distribution $d\sigma/d\log v$\\
{\tt keypre}   &  Defines precision level of the computation\\
               &  For {\tt KeyMod=0} No ISR, Born, in $(e\nu_e)$ channel\\
               &  {\tt =1} absolute error $1\times 10^{-5}$ [pb]\\
               &  {\tt =2} absolute error $1\times 10^{-6}$ [pb]\\
               &  {\tt =3} absolute error $1\times 10^{-7}$ [pb]\\
               &  For {\tt KeyMod>0} ISR, in $(e\nu_e)$ channel\\
               &  {\tt =1} absolute error $3\times 10^{-5}$ [pb]\\
               &  {\tt =2} absolute error $1\times 10^{-5}$ [pb]\\
               &  {\tt =3} absolute error $1\times 10^{-6}$ [pb]\\
               &  {\tt =4} absolute error $1\times 10^{-7}$ [pb]\\
\hline
               &  OUTPUT\\
\hline
{\tt xsect}    &  Cross section  [pb]\\
{\tt errabs}   &  Absolute error [pb]\\
\hline
\end{tabular}
\end{small}
\caption{\it List of arguments of the {\tt korwan} routine. }
\label{tab:korwan-input}
\end{table}

\newpage
\section*{Appendix B:\\ {\tt Umask} matrix for Final States--Default values}

In this appendix we present the template mask for specifying final
states in the {\em fully inclusive} $4$-$fermion$ mode, i.e.\ with 
{\tt KeyWon=KeyZon=1, KeyDwm=KeyDwp=0}.
The mask below opens all channels allowed by the program. 
To exclude some of them, appropriate zeros should be put into the {\tt Umask}
matrix. There is a total of 202 entries in it, corresponding to 
an easy-to-handle classification:
$9\times 9$ of $W\times W$ decay modes and $11\times 11$ of
$Z\times Z$ decay modes. This way some decay channels are
doubly counted as some final states can be both $WW$- and $ZZ$-type. 
To avoid such double-counting in the event generation, appropriate  
entries in the mask are set to $0$. 
Regardless of the input values used, the program
always checks for these MIX-type channels and {\em always} assigns 
them either to $WW$ or to $ZZ$. 
Therefore the user can safely use the mask with all entries set 
to $1$. The actual mask used by the program after this internal
verification is printed to the output file.
Note also that we adopted a convention that 
four of the MIX-type final states 
($u\bar ss\bar u, u\bar bb\bar u, c\bar dd\bar c, c\bar bb\bar c$) are
coded as $ZZ$ ({\tt xpar(1205)}, {\tt xpar(1215)}, {\tt xpar(1227)} and 
{\tt xpar(1229)}) whereas other MIX-types as $WW$ (cf.\ Subsection
\ref{4flib}). Finally, it should be
kept in mind that this is {\em one} mask for {\em all} channels and not
two separate masks for $WW$ and $ZZ$ ones.

{
\small
\begin{verbatim}
xpar(1101-1302)= Umask:

           Wm= 1:ud 2:cd 3:us 4:cs 5:ub 6:cb 7:el 8:mu 9:ta           /Wp=
xpar(1101-1109): 1    1    1    1    1    1    1    1    1              1:ud 
xpar(1110-1118): 1    0    1    1    1    1    1    1    1              2:cd
   ...           1    1    0    1    1    1    1    1    1              3:us 
                 1    1    1    1    1    1    1    1    1              4:cs 
                 1    1    1    1    0    1    1    1    1              5:ub 
                 1    1    1    1    1    0    1    1    1              6:cb 
                 1    1    1    1    1    1    1    1    1              7:el 
                 1    1    1    1    1    1    1    1    1              8:mu 
xpar(1173-1181): 1    1    1    1    1    1    1    1    1              9:ta 
          Z1=  1:d  2:u  3:s  4:c  5:b  6:el 7:mu 8:ta 9:ve 10vm 11vt /Z2=
xpar(1182-1192): 1    0    0    0    0    0    0    0    0    0    0    1:d
xpar(1193-1203): 0    1    0    0    0    0    0    0    0    0    0    2:u 
   ...           1    1    1    0    0    0    0    0    0    0    0    3:s 
                 1    1    0    1    0    0    0    0    0    0    0    4:c 
                 1    1    1    1    1    0    0    0    0    0    0    5:b 
                 1    1    1    1    1    1    0    0    0    0    0    6:el 
                 1    1    1    1    1    1    1    0    0    0    0    7:mu
                 1    1    1    1    1    1    1    1    0    0    0    8:ta
                 1    1    1    1    1    0    1    1    1    0    0    9:ve
                 1    1    1    1    1    1    0    1    1    1    0    10vm 
xpar(1292-1302): 1    1    1    1    1    1    1    0    1    1    1    11vt

\end{verbatim}
}

\newpage
\section*{Appendix C: Output of the Demo Program }
\label{printout}

{
\small
\renewcommand{\baselinestretch}{0.4}
\begin{verbatim}
          *************************************************************** 
          *************************************************************** 
          *************************************************************** 
          *  ###   ###                                   ###       ###  * 
          *  ###  ###  ####    ######      ##     ##     ###       ###  * 
          *  ### ###  ##  ##   ##   ##    ####    ##     ###       ###  * 
          *  ######  ##    ##  ##   ##   ##  ##   ##     ###       ###  * 
          *  ######  ##    ##  #####    ##    ##  ##     ###   #   ###  * 
          *  ### ###  ##  ##   ##  ##   ########  ##      ### ### ###   * 
          *  ###  ###  ####    ##   ##  ##    ##  #######  #### ####    * 
          *  ###   ###           version 1.42.3             ##   ##     * 
          *************************************************************** 
          **********************   March 1999   ************************* 
          *************************************************************** 
                         Last modification:   3.17.1999                   
          *************************************************************** 
          *  Written by:                                                * 
          *    S. Jadach      (Stanislaw.Jadach@cern.ch)                * 
          *    W. Placzek     (Wieslaw.Placzek@cern.ch)                 * 
          *    M. Skrzypek    (Maciej.Skrzypek@cern.ch)                 * 
          *    B.F.L. Ward    (bflw@slac.stanford.edu)                  * 
          *    Z. Was         (Zbigniew.Was@cern.ch)                    * 
          *  Papers:                                                    * 
          *    M. Skrzypek, S. Jadach, W. Placzek, Z. Was               * 
          *      CERN-TH/95-205, Jul 1995, CPC 94 (1996) 216            * 
          *    M. Skrzypek, S. Jadach, M. Martinez, W. Placzek, Z. Was  * 
          *      CERN-TH/95-246, Sep 1995, Phys. Lett. B372 (1996) 289  * 
          *    S. Jadach, W. Placzek, M. Skrzypek, B.F.L. Ward, Z. Was  * 
          *   CERN-TH/98-242, UTHEP-98-0702, Jul 1998, submitted to CPC * 
          *    M. Skrzypek, S. Jadach, W. Placzek, B.F.L. Ward, Z. Was  * 
          *   CERN-TH/99-06, UTHEP-98-1001, Jan 1999, proc. of RADCOR98 * 
          *  Related papers:                                            * 
          *    T. Ishikawa, Y. Kurihara, M. Skrzypek, Z. Was            * 
          *      CERN-TH/97-11, Jan 1997, Eur. Phys. J. C4 (1998) 75    * 
          *    S. Jadach, K. Zalewski                                   * 
          *    CERN-TH/97-29, Jan 1997, Acta Phys. Pol. B28 (1997) 1363 * 
          *  WWW:                                                       * 
          *    http://hpjmiady.ifj.edu.pl/                              * 
          *  Acknowledgements:                                          * 
          *    We acknowledge warmly very useful help of:               * 
          *      M. Martinez in testing versions 1.01 and 1.02,         * 
          *      M. Gruenewald and A. Valassi in testing version 1.21   * 
          *      S. Jezequel in testing versions 1.31-1.33              * 
          *      M. Witek in testing version 1.41                       * 
          *      M. Verzocchi in testing version 1.42                   * 
          *************************************************************** 
                                                                          

 ***************************************************************************
 *                KORALW input parameters used                             *
 *     190.00000000                 CMS energy total         CMSENE    I.0 *
 * ***********************************************                         *
 *             1101                 QED super-switch         KeyRad    IQ1 *
 *                1                 Init. state Rad.         KeyISR    IQ2 *
 *                0                 Final state Rad.         KeyFSR    IQ3 *
 *                1                 Next. To Leading         KeyNLL    IQ4 *
 *                1                 Coulomb corr.            KeyCul    IQ5 *
 * ***********************************************                         *
 *             1012                 Physics super-switc      KeyPhy    IP1 *
 *                0                 FS mass reduction        KeyRed    IP2 *
 *                1                 Spin in W decays         KeySpn    IP3 *
 *                0                 Z propag.                KeyZet    IP4 *
 *                1                 Mass kinematics.         KeyMas    IP5 *
 *                2                 Branching Rat.           KeyBra    IP6 *
 *                0                 W propag.                KeyWu     IP7 *
 * ***********************************************                         *
 *              211                 Technical super-swi      KeyTek    IT1 *
 *                2                 presampler type          KeySmp    IT2 *
 *                1                 rand Numb type           KeyRnd    IT3 *
 *                1                 weighting  switch        KeyWgt    IT4 *
 * ***********************************************                         *
 *            11010                 Miscelaneous             KeyMis    IM1 *
 *                0                 sinW2 input type         KeyMix    IM2 *
 *                1                 4 fermion matr el        Key4f     IM3 *
 *                0                 Anomalous couplings      KeyAcc    IM4 *
 *                1                 WW type final state      KeyWon    IM5 *
 *                1                 ZZ type final state      KeyZon    IM6 *
 * ***********************************************                         *
 *                0                 W-/Z decay mode          KEYDWM    ID1 *
 *                0                 W+/Z decay mode          KEYDWP    ID2 *
 * ***********************************************                         *
 *       1.16639000                 G_mu * 1d5               GMU       I.1 *
 *     128.07000000                 inv alpha_w              ALFWIN    I.2 *
 *      91.18880000                 Z mass   [GeV]           AMAZ      I.3 *
 *       2.49740000                 Z width  [GeV]           GAMMZ     I.4 *
 *      80.23000000                 W mass   [GeV]           AMAW      I.5 *
 *       2.08545732                 W width  [GeV]           GAMMW     I.6 *
 *        .00000100                 dummy infrared cut       VVMIN     I.7 *
 *        .99000000                 v_max ( =1 )             VVMAX     I.8 *
 *       2.00000000                 max wt for rejectn.      WTMAX     I.9 *
 *       4.00000000                 max wt for CC03 rej      WTMAX     I10 *
 *        .12000000                 alpha_s: QCD coupl.      ALPHAS    I11 *
 *        .00000000                 Color Re-Con. Prob.      PReco     I12 *
 * ***********************************************                         *
 *        .23103091                 sin(theta_W)**2          SINW2     I13 *
 * ***********************************************                         *
 * ***********************************************                         *
 *         Z width in Z propagator: s/M_Z *GAMM_Z                          *
 * ***********************************************                         *
 *                                                                         *
 *                            CKM matrix elements:                         *
 *        .97525000                                V_ud   VCKM(1,1)    IV1 *
   ..... skipped ....
 *        .99925000                                V_tb   VCKM(3,3)    IV9 *
 *              Unitarity check of the CKM matrix:                         *
 *                         1.000      .000      .003                       *
 *               VV+ =      .000     1.000      .000                       *
 *                          .003      .000     1.000                       *
 *                                                                         *
 *                   Branching ratios of W decays:                         *
 *        .32097393                                  ud       BR(1)    IB1 *
   ..... skipped ....
 *        .10835195                                 tau       BR(9)    IB9 *
 *                                 fermion masses:                         *
 *        .01000000                                   d   AMAFIN(1)    IM1 *
 *        .00500000                                   u   AMAFIN(2)    IM2 *
   ..... skipped ....
 *       1.77710000                                 tau  AMAFIN(15)   IM10 *
 *        .00100000                                vtau  AMAFIN(16)   IM11 *
 *                                                                         *
 *         Predefined cuts on final state fermions                         *
 *     600.00000000                 min. vis p_t**2           GeV^2     X2 *
 *       8.00000000                add. cut for e+e-ch+       GeV^2     X3 *
 *    .10000000E-05                min. theta with beam        rad      X6 *
 *     300.00000000                max. p_t**2 phot eex       GeV^2     X3 *
 *                                 DECAY LIBRARIES                         *
 *                0                       TAUOLA for W+        JAK1    IL1 *
 *                0                       TAUOLA for W-        JAK2    IL2 *
 *                1                   TAUOLA Ord(alpha)      ITDKRC    IL3 *
 *                1                              PHOTOS      IFPHOT    IL4 *
 *                1                       JETSET for W-      IFHADM    IL5 *
 *                1                       JETSET for W+      IFHADP    IL6 *
 ***************************************************************************

 umask_init=>umask:
  1.0  1.0  1.0  1.0  1.0  1.0  1.0  1.0  1.0
  1.0   .0  1.0  1.0  1.0  1.0  1.0  1.0  1.0
  1.0  1.0   .0  1.0  1.0  1.0  1.0  1.0  1.0
  1.0  1.0  1.0  1.0  1.0  1.0  1.0  1.0  1.0
  1.0  1.0  1.0  1.0   .0  1.0  1.0  1.0  1.0
  1.0  1.0  1.0  1.0  1.0   .0  1.0  1.0  1.0
  1.0  1.0  1.0  1.0  1.0  1.0  1.0  1.0  1.0
  1.0  1.0  1.0  1.0  1.0  1.0  1.0  1.0  1.0
  1.0  1.0  1.0  1.0  1.0  1.0  1.0  1.0  1.0
  1.0   .0   .0   .0   .0   .0   .0   .0   .0   .0   .0
   .0  1.0   .0   .0   .0   .0   .0   .0   .0   .0   .0
  1.0  1.0  1.0   .0   .0   .0   .0   .0   .0   .0   .0
  1.0  1.0   .0  1.0   .0   .0   .0   .0   .0   .0   .0
  1.0  1.0  1.0  1.0  1.0   .0   .0   .0   .0   .0   .0
  1.0  1.0  1.0  1.0  1.0  1.0   .0   .0   .0   .0   .0
  1.0  1.0  1.0  1.0  1.0  1.0  1.0   .0   .0   .0   .0
  1.0  1.0  1.0  1.0  1.0  1.0  1.0  1.0   .0   .0   .0
  1.0  1.0  1.0  1.0  1.0   .0  1.0  1.0  1.0   .0   .0
  1.0  1.0  1.0  1.0  1.0  1.0   .0  1.0  1.0  1.0   .0
  1.0  1.0  1.0  1.0  1.0  1.0  1.0   .0  1.0  1.0  1.0


 ***************************************************************************
 *                         *****TAUOLA LIBRARY: VERSION 2.6 ******         *
 *                         ***********August   1995***************         *
 *                         **AUTHORS: S.JADACH, Z.WAS*************         *
 *                         **R. DECKER, M. JEZABEK, J.H.KUEHN*****         *
 *                         **AVAILABLE FROM: WASM AT CERNVM ******         *
 *                         ***** PUBLISHED IN COMP. PHYS. COMM.***         *
 *                         *******CERN-TH-5856 SEPTEMBER 1990*****         *
 *                         *******CERN-TH-6195 SEPTEMBER 1991*****         *
 *                         *******CERN-TH-6793 NOVEMBER  1992*****         *
 *                         **5 or more pi dec.: precision limited          *
 *                         ******DEXAY ROUTINE: INITIALIZATION****         *
 *                   0     JAK1   = DECAY MODE FERMION1 (TAU+)             *
 *                   0     JAK2   = DECAY MODE FERMION2 (TAU-)             *
 ***************************************************************************

 ***************************************************************************
 *              Window H used only by  Grace 2.0                           *
 *                  Higgs  boson parameters                                *
 *    1000.00000000                xpar(11)= higgs mass         amh     H1 *
 *       1.00000000                xpar(12)= higgs widt         agh     H2 *
 *                                                                         *
 ***************************************************************************

 ***************************************************************************
 *                     Window X_ZZ                                         *
 *                  mm_brancher_ZZ report                                  *
 *                  mm_brancher_ZZ  is on                                  *
 *        .00000000                prob. for branch NR:           1     X1 *
   ..... skipped ....
 *        .02232143                prob. for branch NR:           8     X1 *
   ..... skipped ....
 *        .00000000                prob. for branch NR:          65     X1 *
 *                                                                         *
 ***************************************************************************
\end{verbatim}
}
{
\tiny
\begin{verbatim}
                            Event listing (standard)

    I  particle/jet  K(I,1)   K(I,2) K(I,3)     K(I,4)      K(I,5)       P(I,1)       P(I,2)       P(I,3)       P(I,4)       P(I,5)

    1  !e-!              21       11     0           3           4       .00000       .00000     95.00000     95.00000       .00051
    2  !e+!              21      -11     0           3           4       .00000       .00000    -95.00000     95.00000       .00051
    3  (Z0)              11       23     1           5           6    -41.18296     13.10538     23.92689    129.21950    119.40434
    4  (Z0)              11       23     1           7           8     41.18296    -13.10538    -23.92689     60.78049     35.41166
    5  e-                 1       11     3           0           0       .01444       .00373    -46.54470     46.54470       .00051
    6  e+                 1      -11     3           0           0    -41.19740     13.10166     70.47158     82.67480       .00051
    7  mu-                1       13     4           0           0     -3.64304      1.49896      4.29885      5.83179       .10566
    8  mu+                1      -13     4           0           0     44.82600    -14.60434    -28.22573     54.94870       .10566
                   sum charge:   .00   sum momentum and inv. mass:       .00000       .00000       .00000    190.00000    190.00000
\end{verbatim}
}
{
\small
\renewcommand{\baselinestretch}{0.4}
\begin{verbatim}
 ***************************************************************************
 *                     Window X_WW                                         *
 *                  mm_brancher_WW report                                  *
 *                  mm_brancher_WW  is on                                  *
 *        .46296296                prob. for branch NR:           1     X1 *
   ..... skipped ....
 *        .00000000                prob. for branch NR:          65     X1 *
 ***************************************************************************

 ***************************************************************************
 *                          KORALW  final  report                          *
 *                                Window A                                 *
 *                             WEIGHTED evts.                              *
 *                                                                         *
 *                  ccru matrix element means:                             *
 *     a) Born matrix element for CC03 processes                           *
 *     b) technical crude m.e. for nc processes or                         *
 *        for keysmp .NE. 0                                                *
 *                                                                         *
 *                 xsect with no matrix element                            *
 *            10000                total no of events        nevtot     a0 *
 *                0                wtcrud < 0 evts           nevneg     a1 *
 *    35058.681                    sigma_crude               Xcrude     a2 *
 *    .55820244      +- .64458     <wtcrud>, rel err         wtkacr     a3 *
 *    19569.841      +- 12614.     phsp. vol, no beta-0      xskr       a4 *
 *                                                                         *
 *                                                                         *
 *       xsect with ccru matrix el. only, no betas                         *
 *            10000                total no of events        nevtot     a5 *
 *                0                wtcrud*wtborn <0 evt      nevneg     a6 *
 *    .40047631E-03  +- .27055     <wtcrud*wtborn>, rel      wtkabo     a7 *
 *    14.040171      +- 3.7986     sigma (born m.el.)         xska0     a8 *
 *                                                                         *
 *                      xsect over wtmax_cc03                              *
 *                  ccru matrix el. only, no betas                         *
 *                0                evts: wt>wtmax_cc03       nevove     a9 *
 *    .00000000E+00  +- .00000E+00 sigma: wt>wtmax_cc03      xskabo    a10 *
 *    .00000000E+00  +- .00000E+00 relat sigma: wt>wtma      xskabo    a11 *
 ***************************************************************************

 ***************************************************************************
 *                          KORALW  final  report                          *
 *                                Window B                                 *
 *                             Xsec-s in [pb]                              *
 *                                                                         *
 *      29.68888933  +- 4.74480757       xsec total         O(alf0)     b3 *
 *      30.29512756  +- 4.95372145       xsec total         O(alf1)     b4 *
 *      30.33902996  +- 4.96038178       xsec total         O(alf2)     b5 *
 *      30.33978251  +- 4.96052513       xsec total         O(alf3)     b6 *
 *                0                      wt<0  events       O(alf0)        *
 *                0                      wt<0  events       O(alf1)        *
 *                0                      wt<0  events       O(alf2)        *
 *                0                      wt<0  events       O(alf3)        *
 *      29.68888933  +- 4.74480757       xsec(beta00)       O(alf0)     b7 *
 *      31.38896196  +- 5.01650913       xsec(beta01)       O(alf1)     b8 *
   ..... skipped ....
 *        .02364120  +-  .01578300       xsec(beta21)       O(alf3)    b15 *
 *       -.00003663  +- -.00003099       xsec(beta30)       O(alf3)    b16 *
 *                           xsec_tot differences                          *
 *        .60623824  +-  .29779330        xstot(alf1-0)     O(alf1)    b17 *
 *        .04390240  +-  .01499107        xstot(alf2-1)     O(alf2)    b18 *
 *        .00075255  +-  .00022070        xstot(alf3-2)     O(alf3)    b19 *
 *                              betas differences                          *
 *       1.70007263  +-  .27170156        xs(beta01-00)     O(alf1)    b20 *
   ..... skipped ....
 *       -.00055056  +- -.00019179        xs(beta21-20)     O(alf3)    b24 *
 *       -.00003663  +- -.00003099        xs(beta30)        O(alf3)    b25 *
 ***************************************************************************

 ***************************************************************************
 *                          KORALW  final  report                          *
 *                                Window C                                 *
 *                                                                         *
 *                      BEST order total xsect.                            *
 *            10000                 total no of events       nevtot     c1 *
 *            10000                accepted events           NevTru     c2 *
 *    30.339783      +- 4.9605      sigma_tot [pb]           xskabo     c3 *
 *        .16349903                 relative error           errela     c4 *
 *                0                 events: wt<0             nevneg     c5 *
 *    .00000000E+00  +- .00000E+00  xsec/xtot: wt<0          xsneg      c6 *
 *                0                 events: wt>wtmax         nevove     c7 *
 *    .00000000E+00  +- .00000E+00  xsec/xtot: wt>wtmax      xsove      c8 *
 ***************************************************************************

 ***************************************************************************
 *                          KORALW  final  report                          *
 *                                Window D                                 *
 *                                                                         *
 *                      Complete 4-fermion process                         *
 *                                                                         *
 *            I. Best ord. W-pair total xsect.                             *
 *    .41443001E-03  +- .27359      <wttww>: WW weight       averwt     d1 *
 *    14.529370      +- 3.9751      sigma_WW, best [pb]      xskabo     d2 *
 *                                                                         *
 *            II. Best ord. 4-fermion total xsect.                         *
 *    21819.562      +- .74956      <wtbo4f>, rel err        averwt     d3 *
 *    .86540000E-03  +- .16350      <wttot>,rel err          averwt     d4 *
 *    30.339783      +- 4.9605      sigma_4f, best [pb]      xskabo     d5 *
 *    .52111161      +- .15263      sigma 1-Wpair/4ferm      1-d2/5     d6 *
 *    .52111161      +- .12996      sigma 1-Wpair/4ferm      wtbgr      d7 *
 ***************************************************************************
 
 ***************************************************************************
                      Decay Report on Different Channels 
 
                                             wt_max  wt_max  nev_ch nev_non0
  wm wp        human   sigma [pb] +- abs_err ------ -------- ------ --------
                                              <wt>  <wt_non0>   tot   nev_ch
 
   1  1  dq~uq uq~dq  .3863394E+01+- .25E+01  .67E+01  .37E+01  .0011  .5455  
   2  1  dq~cq uq~dq  .0000000E+00+- .00E+00  .67E+01  .37E+01  .0000  .0000
   3  1  sq~uq uq~dq  .0000000E+00+- .00E+00  .67E+01  .37E+01  .0000  .0000  
   4  1  sq~cq uq~dq  .6681917E-09+- .67E-09  .20E+01  .10E+01  .0002  .5000
   ..... skipped ....
  11 11  nt~nt nt~nt  .0000000E+00+- .00E+00  .10E+01  .10E+01  .0000  .0000
  
    total xsection =  30.3397825138152157  +-  4.96052512935006362  [pb]
 ***************************************************************************
  
                              ============ demo ============
       30.33978251  +- 4.96052513          MC Best, XPAR, KoralW
                              ========== End demo ==========
\end{verbatim}
}

\newpage
\bibliographystyle{prsty}


\end{document}